\documentclass[a4paper]{ar-1col}

\usepackage{natbib}
\usepackage{amssymb,amsmath}
\usepackage{xspace}
\usepackage[usenames,dvipsnames]{xcolor}
\usepackage[backref,
            breaklinks,
            colorlinks,
            urlcolor=Red,
            linkcolor=Maroon,
            citecolor=Blue]{hyperref}
\usepackage[capitalise,nameinlink]{cleveref}
\usepackage[all]{hypcap}

\jname{Annu. Rev. Astron. Astrophys.}
\jvol{58}
\jyear{2020}
\jnote{submitted 2019/09/15; accepted 2019/11/15.  An edited final version will be published later in 2020}
\begin{document}

\markboth{Andrews}{Disk Structures}

\title{Observations of Protoplanetary Disk Structures}

\author{Sean M. Andrews
\affil{Center for Astrophysics \textbar\ Harvard \& Smithsonian \\ Cambridge, Massachusetts, USA 02138; email: sandrews@cfa.harvard.edu}}

\begin{abstract}
The disks that orbit young stars are the essential conduits and reservoirs of material for star and planet formation.  Their structures, meaning the spatial variations of the disk physical conditions, reflect the underlying mechanisms that drive those formation processes.  Observations of the solids and gas in these disks, particularly at high resolution, provide fundamental insights on their mass distributions, dynamical states, and evolutionary behaviors.  Over the past decade, rapid developments in these areas have largely been driven by observations with the Atacama Large Millimeter/submillimeter Array (ALMA). This review highlights the state of observational research on disk structures, emphasizing three key conclusions that reflect the main branches of the field: 
\vspace{0.2cm}
\\
$\bullet$ Relationships among disk structure properties are also linked to the masses, environments, and evolutionary states of their stellar hosts;  \vspace{0.1cm} \\
$\bullet$ There is clear, qualitative evidence for the growth and migration of disk solids, although the implied evolutionary timescales suggest the classical assumption of a smooth gas disk is inappropriate; \vspace{0.1cm} \\
$\bullet$ Small-scale substructures with a variety of morphologies, locations, scales, and amplitudes -- presumably tracing local gas pressure maxima -- broadly influence the physical and observational properties of disks.  
\vspace{0.2cm}
\\
The last point especially is reshaping the field, with the recognition that these disk substructures likely trace active sites of planetesimal growth or are the hallmarks of planetary systems at their formation epoch.                  
\end{abstract}


\maketitle

\tableofcontents

\section{INTRODUCTION} \label{sec:intro}

\subsection{Motivation} \label{sec:motivation}

The formation and early evolution of stars and planetary systems are mediated by interactions with their circumstellar material.  That material is organized in a flattened disk of gas and solids that orbits the central host star.  Although these interactions between stars, planets, and disks are brief (lasting $\lesssim$\,10 Myr), they are literally foundational: such mutual influences set some stellar and planetary properties that persist for billions of years.  The hallmarks of the processes that govern these links are imprinted on the disk {\it structures}, the spatial distributions and physical conditions of the disk material.  Detailed observations enable measurements of those structures, their environmental dependencies, and their evolutionary behavior.  Coupled with theoretical simulations and complemented by the collective knowledge of stellar populations, exoplanets, and primitive bodies in the solar system, those measurements help map out how disks shape star and planet formation.   

These disks and their initial structures are seeded when a star is made.  Star formation begins with the gravitational collapse of an over-dense core in a molecular cloud.  An initial nudge that imparts some core rotation means that material collapsing from its outer regions (with higher angular momentum) is channeled onto a disk, rather than the protostar itself \citep{terebey84}.  In that sense, disks are simple consequences of angular momentum conservation.  Measurements of young disk structures, still embedded in their natal core material, can reveal much about the star formation process: their sizes help distinguish the roles that magnetic fields have in regulating core collapse; their masses help constrain protostellar accretion rates; and their density distributions encode the angular momentum transport that ultimately determines the stellar mass (see the review by \citealt{li14}).  

Disks are also the material reservoirs and birthplaces of planetary systems.  The prevalence, formation modes, masses, orbital architectures, and compositions of planets depend intimately on the physical conditions in the disk at their formation sites, the evolution of that disk structure (locally and globally), and the planetary migration driven by dynamical interactions with the disk material.  Measurements of the disk mass, its spatial distribution, and its demographic dependences offer crucial boundary conditions for models of planet formation.  Combined with the properties observed in the mature exoplanet population, that information can help develop and refine a predictive formation theory, despite the considerable complexity of the associated physical processes \citep[e.g.,][]{benz14}.

\begin{figure}[t]
\includegraphics[width=\textwidth]{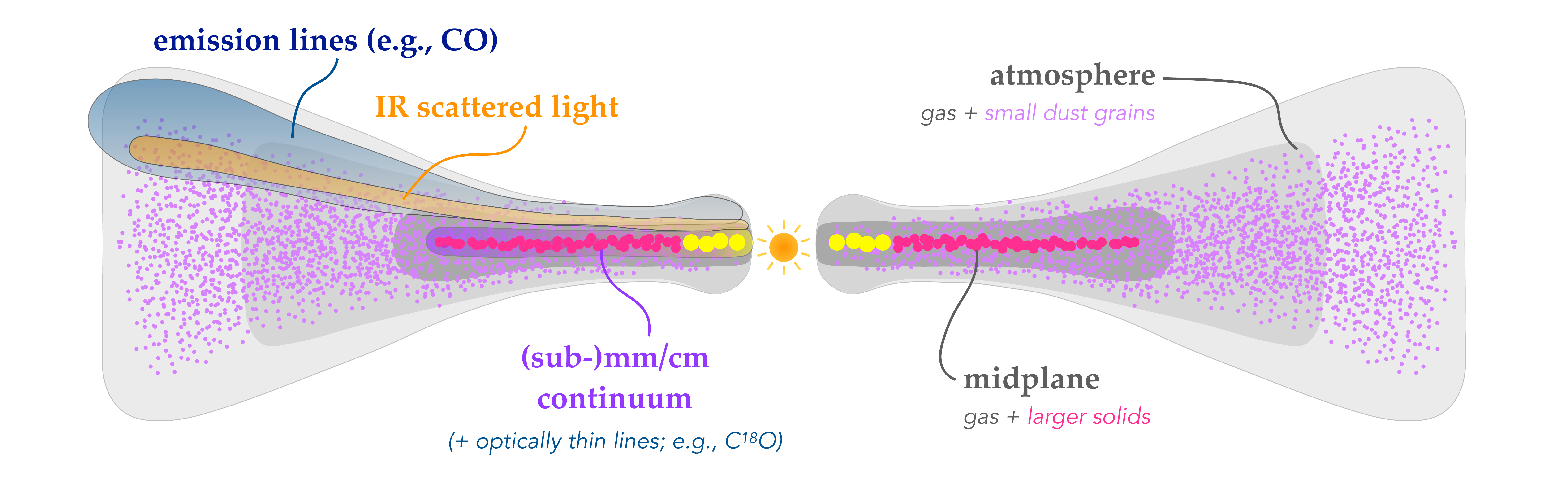}
\caption{A cartoon schematic of a disk structure viewed in cross-section.  The gas is denoted in grayscale, and solids are marked with exaggerated sizes and colors.  The left side highlights the approximate locations of emission tracers; the right side defines some structure and contents terminology.}
\label{fig:cartoon_overview}
\end{figure}

\subsection{Observational Primer} \label{sec:primer}

In these and many other ways, disk structures offer profound insights on how the properties of stars and planetary systems are shaped by their origins.  This review is focused on the recent landscape of observational constraints on disk structures: how relevant measurements are made, what they suggest about disk properties, and how those properties are connected to star and planet formation.  The most valuable measurements employ data with high angular resolution, as the typical nearby ($d \approx 150$ pc) disk subtends $\lesssim$\,1$^{\prime\prime}$ on the sky.  Most of any given disk is cool enough ($<$\,100 K) that it emits efficiently at (sub-)mm wavelengths.  Coupling these small angular sizes and cool temperatures, this review emphasizes radio interferometry as an essential tool.  Indeed, progress over the past decade has largely been driven by the commissioning of the transformational ALMA facility.

\begin{figure}[t]
\includegraphics[width=\textwidth]{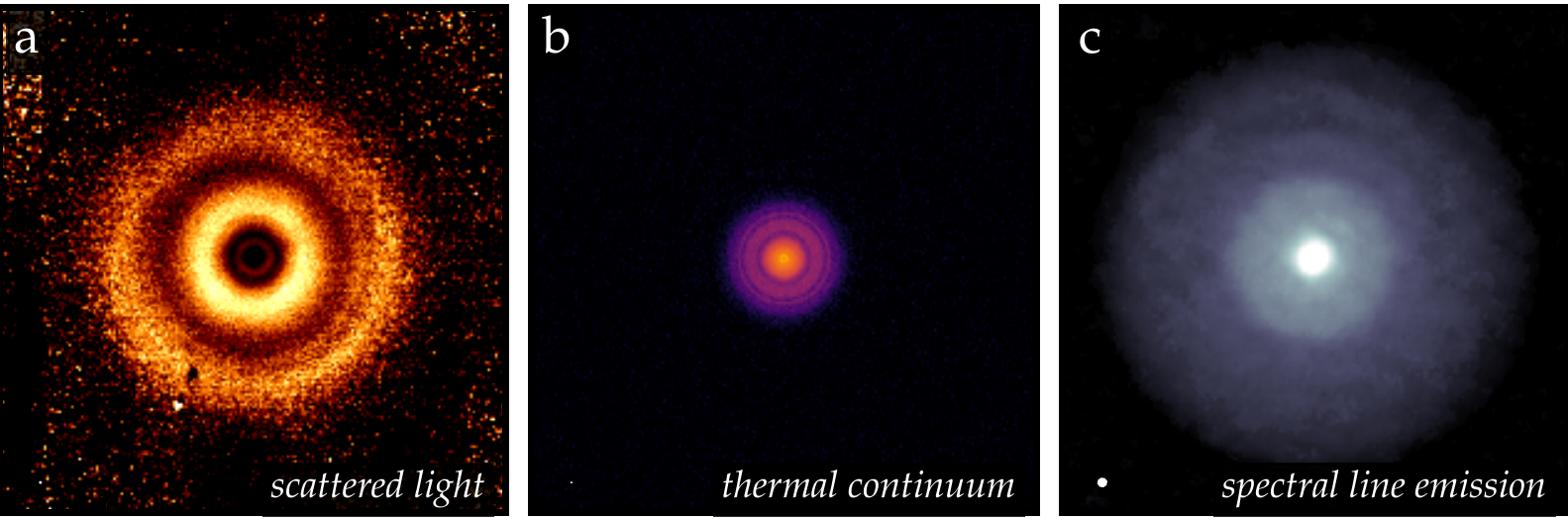}
\caption{The morphology of the TW Hya disk is compared in three different tracers: (a) $\lambda = 1.6$ $\mu$m scattered light from small dust grains \citep{vanboekel17}, (b) $\lambda = 0.9$ mm continuum from pebble-sized particles \citep{andrews16}, and (c) the CO $J$=3$-$2 spectral line emission tracing the molecular gas \citep{huang18}.  Each panel spans 500 au on a side; resolutions are shown with ellipses in the lower left corner of each panel (too small to be visible in a and b).  It is helpful to compare these emission distributions with the behavior in the {\bf Figure \ref{fig:cartoon_overview}} schematic.  }
\label{fig:ims}
\end{figure}

Three categories of observational tracers are used to study disk structures: scattered light, thermal continuum emission, and (primarily molecular) spectral line emission.  The first two are sensitive to the physical conditions and distribution of the solids, and the third is used to measure the properties of the gas.  {\bf Figure \ref{fig:cartoon_overview}} shows a schematic diagram that highlights the basic aspects of disk structure and the (two-dimensional) locations where these tracers originate.  Each of these probes is sensitive to different materials and physical conditions, ensuring considerable diversity in the disk appearance when viewed in different tracers.  An illustrative example is shown in {\bf Figure \ref{fig:ims}}.

\subsubsection{Scattered Light} \label{sec:primer_scat}
Small ($\sim$$\mu$m-sized) dust grains suspended in the gas at a suitable altitude in the disk atmosphere reflect the radiation emitted by the host star.  This scattered light is sensitive to the radial variation of the vertical height of the dust distribution (Section \ref{sec:dustevol}).  The spectral and polarization behavior of the scattered light constrain the albedos, set by the sizes, shapes, and compositions of the grains (Section \ref{sec:opacities}).  The practical advantage of this tracer is resolution: adaptive optics systems operating near the diffraction limit on 8--10 m telescopes measure features at 30--50 milliarcsecond scales ($\sim$5 au at the typical distances of nearby star-forming regions, $\sim$150 pc).  The important challenges include: contrast with the host star, preventing measurements in the innermost disk ($\lesssim$\,10 au); sensitivity at large radii, due to the dilution of the stellar radiation field; and technical limits on the host star brightness.  Taken together, those issues bias the current sample of resolved scattered light measurements toward disks with more massive hosts.

\subsubsection{Continuum Emission} \label{sec:primer_cont}
Disk solids emit a thermal continuum that spans four decades in wavelength ($\lambda \approx 1\;\mu$m--1 cm).  Most of that emission is optically thick, and therefore a temperature diagnostic.  Optical depths ($\tau_\nu$) decrease with $\lambda$; the transition to $\tau_\nu \lesssim 1$ is traditionally expected in the sub-mm.  In the optically thin limit, the intensity ($I_\nu$) scales with the surface density of solids ($\Sigma_{\rm s}$; Section \ref{sec:mass}, \ref{sec:sigma}), and its spectral dependence is sensitive to the solid particle properties (Section \ref{sec:dustevol}).  This tracer is bright, accessible at high resolution (to 10--20 milliarcseconds, or $\sim$2 au), and has no stellar contrast limitations.  Accordingly, measurements are plentiful: much of the collective knowledge about disk structures is based on mm continuum data.  The disadvantages arise from ambiguities in the detailed particle properties and the validity of the optically thin approximation (Sections \ref{sec:solids} and \ref{sec:substructures}).

\subsubsection{Spectral Line Emission} \label{sec:primer_lines}
The most abundant molecule in a disk (H$_2$) does not have a permanent dipole moment and does not emit efficiently over the vast majority of the disk volume.  The bulk of the gas in a disk is essentially `dark', and there is no {\it direct} probe of its mass reservoir.  Instead, measurements rely on the spectral line emission from (sub-)mm rotational transitions of rare tracer molecules.  Optically thick line intensities are sensitive to the temperature in the atmospheric layer that corresponds to the line photosphere (Section \ref{sec:temp}).  At low optical depths, line intensities are a function of both temperature and density.  If the abundance of a given species relative to H$_2$ (denoted here as {\sf X}$_j$ for molecule $j$) is known, spatially resolved maps of optically thin line emission constrain the gas surface density profile, $\Sigma_{\rm g}$ (Section \ref{sec:mass}, \ref{sec:sigma}).  Moreover, spectrally resolved line emission can be used to tomographically reconstruct the disk velocity field (Section \ref{sec:turb}).  

ALMA is now capable of resolving emission lines at tens of milliarcsecond scales ($\sim$5 au) in velocity channels only a few m s$^{-1}$ wide, but the narrow bandwidths and low abundances of trace molecular species mean that sensitivity is a perennial challenge for disk measurements.  Accordingly, line measurements of disks are much less common than for the continuum.  The most prominent obstacles in interpreting spectral line data are high optical depths, confusion with the emitting layer height (when resolution is limited), and the large (potentially orders of magnitude) uncertainties in the molecular abundances ({\sf X}$_j$).

\subsection{Statement of Scope} \label{sec:scope}

Keeping in mind the motivations for measuring disk structures and the observational tools that are now available, this review covers four broad (and inter-related) topics that occupy much of the effort in the disk research community: inferred physical characteristics of disk structures and their ambiguities (Section \ref{sec:structure}); empirical constraints on evolutionary and environmental dependencies based on demographics studies (Section \ref{sec:demographics}); evidence for (and problems with) the growth and migration of disk solids (Section \ref{sec:solids}); and the properties and roles of small-scale {\it substructures} in shaping observables and facilitating planet formation (Section \ref{sec:substructures}).  The review concludes with a brief synopsis that summarizes the current state of the field and some suggestions of potentially fruitful avenues for future work (Section \ref{sec:summary}).

\section{KEY\ STRUCTURE\ PROPERTIES} \label{sec:structure}

The spatial distribution of mass -- the density structure -- is without question the fundamental property of interest for disks.  The conceptual orientation of the entire field presumes that disk evolution is deterministic: in principle, a collection of density structure measurements that span an appropriate range of environmental and evolutionary states could be used to work out the mechanics of key evolutionary processes.  This section of the review is focused on the underlying motivations, observational constraints, and lingering ambiguities associated with the mass distributions in disks (Sections \ref{sec:mass}--\ref{sec:sigma}).  The intrinsic connections (physical and observational) between the density structure and the thermal and dynamical state of the disk material are summarized in Sections \ref{sec:temp} and \ref{sec:turb}, respectively.

\begin{textbox}[t]\section{Notation, Conventions, and Nomenclature}
To simplify discussions, variable notations are used throughout this review.  Cylindrical coordinates define the disk frame of reference for these properties, where ($r$, $\theta$, $z$) correspond to the radial distance from the star, the azimuthal angle around the disk, and the height above the disk midplane, respectively.  Many structural and observational properties vary in three dimensions.  To minimize complexity and avoid confusion, the convention is to explicitly note spatial dependencies only when the spatial behavior is directly relevant (e.g., most discussion presumes azimuthal symmetry).  For example: $T$ is shorthand for the local value $T(r, \theta, z$); $T(r)$ refers to a radial profile at a given (e.g., the midplane) or generic $z$, depending on the context; and $\langle T \rangle$ refers to a disk-averaged quantity, $\langle T \rangle = \iiint T(r, \theta, z) \, r \, dr \, d\theta \, dz / \! \iiint  r \, dr \, d\theta \, dz$.
\end{textbox}

\subsection{Mass} \label{sec:mass}

With a limited number of resolved disk measurements, more emphasis is placed on masses than density distributions.  Nevertheless, the key issues can be illustrated from this coarser perspective.  Masses offer elementary constraints on the future contents of planetary systems.  The current census of exoplanets finds an abundance of worlds orbiting other stars, but the metamorphosis of disk material into planetary systems is unclear without a comparison of the available mass reservoirs in the parent and descendant populations.  Summing the masses of terrestrial planets and giant planet cores in the solar system, or an ensemble of exoplanets, offers a conservative lower bound on the solid mass expected in their progenitor disks, $M_{\rm s} \gtrsim 40$ M$_\oplus$ \citep{weidenschilling77b,chiang13}.  Extrapolations of the current planetary atmosphere compositions to the primordial gas expected in disks give an analogous bound for the gas masses, $M_{\rm g} \gtrsim 3000$ M$_\oplus$.

\begin{figure}[t]
\includegraphics[width=\textwidth]{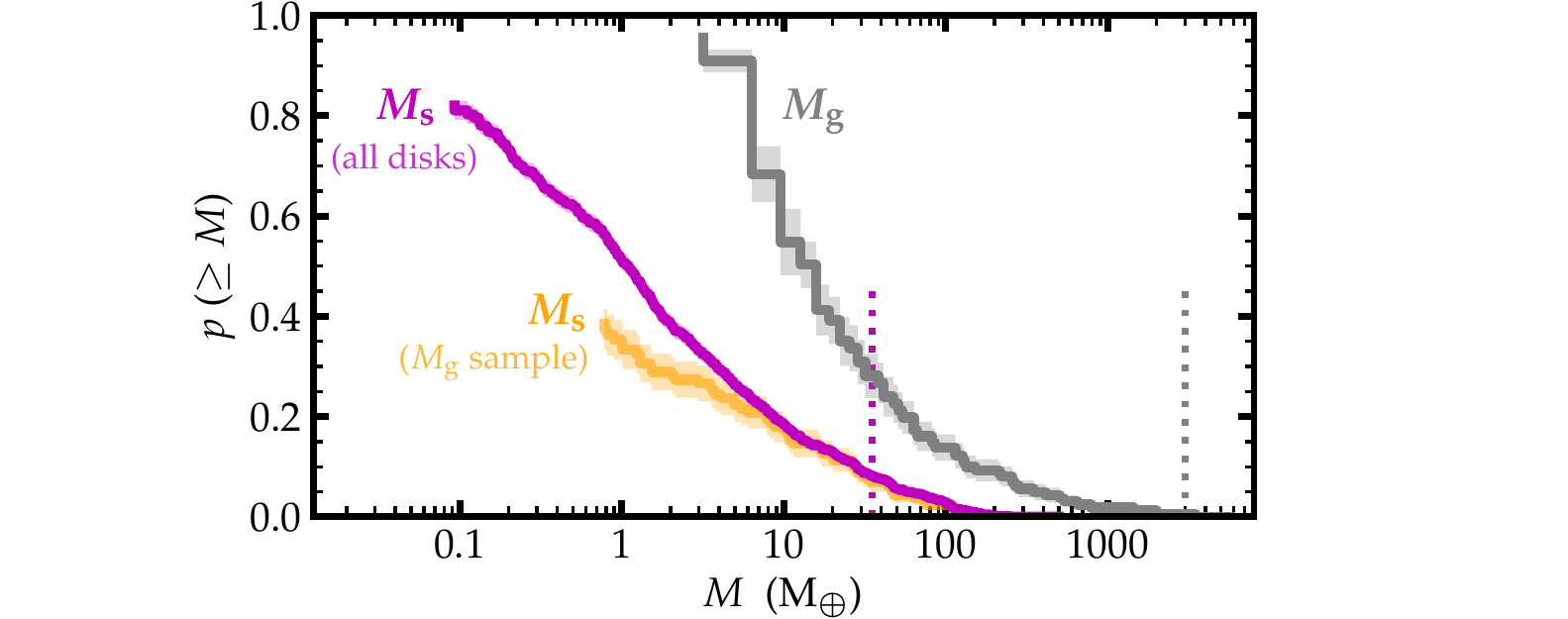}
\caption{Cumulative distributions of disk solid ($M_{\rm s}$) and gas ($M_{\rm g}$) masses.  The $M_{\rm s}$ distribution (purple) includes 887  disks in Oph \citep{cieza19,williams19}, Tau \citep{andrews13,akeson19}, Lup \citep{ansdell16,ansdell18}, Cha \citep{pascucci16,long18b}, IC 348 \citep{ruiz-rodriguez18}, and Upper Sco \citep{barenfeld17}.  $L_{\rm mm}$ measurements at $\lambda = 0.9$ mm (or 1.3 mm, scaled by $\nu^{2.2}$; Section \ref{sec:growth-mm}) were converted to $M_{\rm s}$ assuming $\langle T \rangle = 20$ K and $\langle \kappa_\nu \rangle = 3.5$ cm$^2$ g$^{-1}$.  The $M_{\rm g}$ distribution (gray) uses CO isotopologue data (\citealt{ansdell18,long18b}; with a small supplement from individual case studies) and employs the models of \citet{miotello17}.  Since such line data are rare and the sample biased, the $M_{\rm s}$ distribution for the same disks is also shown (orange) for comparison.  The minimum masses in solids and gas needed to produce the solar system planets are shown as dashed vertical lines in purple and gray, respectively (following \citealt{weidenschilling77b}).}
\label{fig:mass_dist}
\end{figure}

Solids are a minor contributor to the mass budget, with an initial mass fraction of $\sim$1\%\ relative to the gas.  But the fundamental roles they play in all aspects of disk evolution and planet formation justify special attention to their mass reservoir.  The optimal $M_{\rm s}$ diagnostic is the luminosity of the mm continuum emission, $L_{\rm mm}$.  In the optically thin limit, the continuum intensity scales like $I_\nu \propto \kappa_\nu \, B_\nu(T) \, \Sigma_{\rm s}$, where $\kappa_\nu$ is the absorption opacity, $B_\nu(T)$ the Planck function at temperature $T$, and $\Sigma_{\rm s}$ the surface density of solids.  Integrating that emission over the disk volume shows that $L_{\rm mm} \propto M_{\rm s}$.  {\bf Figure \ref{fig:mass_dist}} shows the $M_{\rm s}$ distribution inferred from mm continuum photometry surveys for 887 disks.  Though that distribution is subject to considerable ambiguities (see below) and biased by observational and physical selection effects (Section \ref{sec:demographics}), it offers rough guidance on typical $M_{\rm s}$ values.  

Estimates of $M_{\rm s}$ are intrinsically uncertain because they rely on assumptions about the properties of the emitting particles.  A detailed discussion of those properties, encapsulated in the absorption opacity, is deferred to Section \ref{sec:solids}, but the standard approach is to adopt a reasonable estimate that maximizes $\langle \kappa_\nu \rangle$.  Coupled with the possibility that some of the continuum emission is optically thick \citep{beckwith90,aw05,zhu19}, this implies that $M_{\rm s}$ estimates are more appropriately considered {\it lower bounds}.  The sense of that ambiguity factors into comparisons between the distributions of disk and planetary system masses, as highlighted in the box at the end of this section.  

There are many fewer estimates of $M_{\rm g}$, primarily because mm spectral line observations are more expensive than for the continuum.  One option for a mass-sensitive tracer molecule is HD, the primary isotopologue of H$_2$ \citep{bergin13,mcclure16}.  The advantage of HD is the simplicity of its associated chemical network, which builds confidence in estimates of its abundance, {\sf X}$_{\rm HD}$.  But with a ground state transition at 112 $\mu$m, HD measurements are scarce (three disk detections) and currently inaccessible (with no operational far-infrared space telescope).  Estimates of $M_{\rm g}$ based on HD have a strong $T$-dependence and are considered lower bounds for two reasons.  First, there are potential alternative reaction pathways (e.g., into hydrocarbons) that could lower {\sf X}$_{\rm HD}$.  And second, the line may be optically thick, and some of the emission could be hidden below the optically thick local continuum.  These latter issues can be treated by comparing the data with radiative transfer models that interpret a prescription for the two-dimensional temperature and density structures \citep[e.g.,][]{mcclure16,trapman17}.   

CO is a more common gas tracer in disks, since the abundance is high and the low-energy rotational transitions are easily accessed with mm interferometers.  The primary isotopologue has very high optical depths \citep{beckwith93}, so $M_{\rm g}$ estimates rely instead on rarer species (usually $^{13}$CO and C$^{18}$O together) and references to parametric model catalogs \citep{williamsbest14,miotello16}.  A modest (and biased) collection of $M_{\rm g}$ measurements are available from assorted case studies and shallow line surveys \citep[e.g.,][]{ansdell16,ansdell18,long17}, as shown in {\bf Figure \ref{fig:mass_dist}}.  

\begin{textbox}[t]\section{Is there enough mass in disks to make planetary systems?}
The mass distributions in {\bf Figure \ref{fig:mass_dist}} suggest that few disks have enough material to produce the solar system or its counterparts in the exoplanet population.  Interpretations of this discrepancy have been considered in various forms \citep{greaves10,najita14,manara18}, with proposed solutions falling into two (not mutually exclusive) categories.  The first explanation is perhaps pessimistic, but it simply recalls that $M_{\rm s}$ and $M_{\rm g}$ estimates are {\it lower bounds}: biased accounting factors ($\kappa_\nu$ or {\sf X}$_j$) and optically thick contamination could make the true masses much higher.  For example, if the mm continuum emission used to estimate $M_{\rm s}$ includes contributions from 10 cm rocks instead of 1 mm pebbles, the $M_{\rm s}$ distribution would shift up an order of magnitude (Section \ref{sec:opacities}).  The second solution strikes a more optimistic tone, proposing instead that planet formation has already occurred and the observations are tracing the ``leftovers" (e.g., collisional debris) rather than the actual mass.  The concept of a condensed planet formation timescale, presumably occurring during the embedded phase \citep[e.g.,][]{nixon18}, has gained recent momentum from the fine-scale features that are now routinely identified in high resolution disk images (Section \ref{sec:substructures}).    
\end{textbox}

These CO-based masses appear low, $\sim$5--10$\times$ lower than crude estimates from the product of the accretion rate and stellar age \citep{manara16}, or different gas tracers in the same disks \citep{favre13,kama16a}, or if $M_{\rm s}$ is scaled up by a standard gas-to-solids fraction (100; \citealt{dutrey03,ansdell16}).  The anomaly can be reconciled with a lower gas-to-solids ratio ($\lesssim$\,10) or by decreasing {\sf X}$_{\rm CO}$ or the isotope fractionation below ISM values.  Such abundance changes are expected from various processes \citep[e.g.,][]{miotello17}, including adsorption onto solids (\citealt{aikawa97,vanzadelhoff01}), isotope-selective photodissociation \citep{miotello14,miotello16}, and especially the sequestration of C or O into grains, ices, or other species (e.g., organics; \citealt{reboussin15,yu16,yu17a,miotello17,bosman18}).  Alternatively, the typical CO isotopologue tracers (even C$^{18}$O) might be optically thick, saturating the line luminosities \citep{booth19}.  The salient point is again that the standard adopted assumptions produce lower bounds on $M_{\rm g}$ by design.
\begin{marginnote}[]
\entry{ISM}{interstellar medium.}
\end{marginnote}

\subsection{Size} \label{sec:sizes}

Sizes are a natural step in the progression of measurements from masses to density profiles.  There is no consensus size definition, physically or observationally, since any metric depends on the adopted prescription for the radial variations of densities or intensities.  A physical modeling effort to homogenize size measurements is littered with ambiguities.  A more practical approach is to assign an empirical definition of an effective size, $R_j$, defined as the radius that encircles a fixed fraction of the luminosity from tracer $j$. 

\begin{figure}[t]
\includegraphics[width=\textwidth]{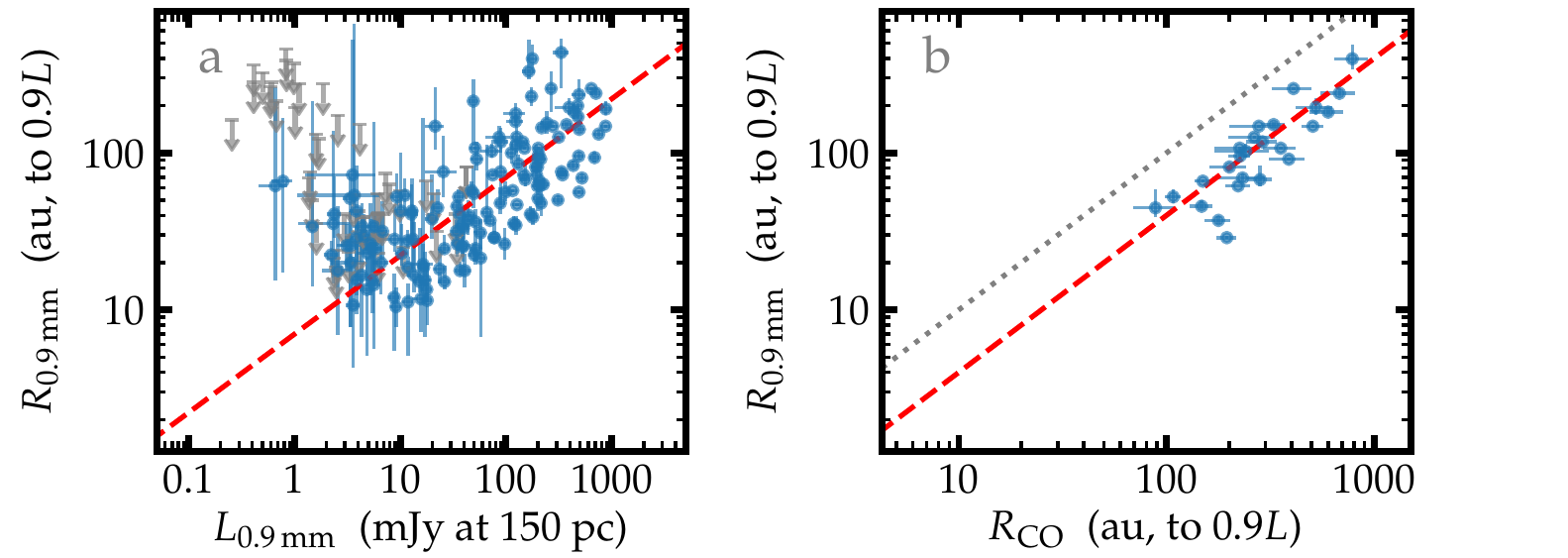}
\caption{(a) The correlation between the $\lambda = 0.9$ mm continuum luminosities (in flux density units, scaled to a common distance of 150 pc) and sizes (the radii that encircle $0.9\,L_{\rm mm}$; \citealt{tripathi17,andrews18,hendler19}).  The inferred scaling relation, $R_{\rm eff} \propto L_{\rm mm}^{0.5}$, is overlaid in red.  (b) A comparison of $R_{\rm mm}$ and analogous sizes inferred from the CO line emission (data from \citealt{oberg11,msimon17,ansdell18,facchini19}).  A one-to-one marker is shown as a dotted line, and a $R_{\rm CO} \approx 2.5 \, R_{\rm mm}$ relation is shown in red.}
\label{fig:sizes}
\end{figure}

Resolved mm continuum measurements from roughly 200 disks have been used to infer $R_{\rm mm} \approx 10$--500 au (defined here so $R_{\rm mm}$ encircles 0.9\,$L_{\rm mm}$; \citealt{tripathi17,andrews18,hendler19}).  The lower bound of that range is presumably limited by resolution.  {\bf Figure \ref{fig:sizes}a} shows a tight correlation between the mm continuum sizes and luminosities \citep{andrews10,pietu14} with a scaling relation $L_{\rm mm} \propto R_{\rm mm}^2$ \citep{tripathi17,andrews18} that may flatten for older systems \citep{hendler19}.  The origins of this relationship are not clear: it could be imposed at the disk formation epoch \citep{isella09,andrews10}, produced by the evolution of solids (\citealt{tripathi17,rosotti19b}; see Section \ref{sec:solids}), or it may be a more trivial manifestation of high optical depths \citep{andrews18,zhu19}.  

Scattered light images offer an alternative size metric for the solids, although the empirical methodology outlined above has not been used for such data in the literature.  Nevertheless, the current suite of scattered light images \citep[e.g.,][]{garufi18} demonstrate that the $\mu$m-sized dust grains that reflect starlight are distributed out to greater distances than the larger particles responsible for the mm continuum (e.g., {\bf Figure \ref{fig:ims}}).  Quantifying this size difference can be especially difficult, in large part due to the dilution of the stellar radiation field in the outer disk, but future comparisons would be valuable.

Again, there are many fewer size measurements for the gas phase.  The CO line emission extends to $R_{\rm CO} \approx 100$--500 au \citep{ansdell18}, although a few outliers stretch beyond the high end of that range.  Smaller disks presumably exist, but produce such weak line emission that they are missing in current samples \citep[e.g.,][]{barenfeld17_sizes}.  The available data suggest $R_{\rm CO} \gtrsim 2 \, R_{\rm mm}$, as shown in {\bf Figure \ref{fig:sizes}b}.  Some of that difference is related to comparing tracers with such different optical depths \citep{hughes08,trapman19}, but radiative transfer models argue for a genuine discrepancy between the density distributions  \citep{panic09,andrews12,facchini19}.  

There is a significant caveat in these measurements and results that merits reiteration: these are {\it empirical} size measurements that are not directly or simply linked to the density distribution.  While the inferred behaviors outlined above may point to fundamental physical relationships, a translation into physical radii is not obvious \citep{rosotti19a}.

\subsection{Density} \label{sec:sigma}

In principle, spatially resolved measurements of the mass tracers introduced in Section \ref{sec:mass} can constrain the surface density profiles for the solids ($\Sigma_{\rm s}$) or gas ($\Sigma_{\rm g}$).  Measurements of $\Sigma_{\rm g}$ offer important insights on how angular momentum is transported in disks through turbulent viscosity \citep{hartmann98} or winds \citep{blandford82,bai13}, and what types of planetary system architectures can be formed \citep{miguel11} and how they will evolve via migration \citep{baruteau14}.  Likewise, the evolution of $\Sigma_{\rm s}$ is a diagnostic of the processes that drive the growth of dust grains into planetesimals \citep{johansen09,birnstiel12}.  Put simply, the disk density structure ties into all the fundamental physical mechanisms relevant to star and planet formation.        

A key emphasis has been on estimating $\Sigma_{\rm s}$ from modest resolution ($\sim$20--50 au) observations of mm continuum morphologies \citep{aw07a,pietu07,pietu14,andrews09,andrews10,isella09,tazzari17}.  The modeling details used to make those measurements vary substantially between studies, but a crude distillation suggests $\Sigma_{\rm s} \propto r^{-1}$ or shallower in the inner disk and $\Sigma_{\rm s} \propto r^{-3}$ or steeper at large $r$.  At these resolutions, the density gradient is actually measured at large $r$ (tens to hundreds of au); estimates of $\Sigma_{\rm s}$ in the inner disk are extrapolated according to the prescribed functional form of the density profile.  At $r \approx 50$--100 au,  $\Sigma_{\rm s}$ values span $\sim$0.001--1 g cm$^{-2}$.

The basic methodology for inferring $\Sigma_{\rm g}$ from spectral line observations is similar \citep[e.g.,][]{williams16,zhang19}.  The focus has been almost exclusively on the CO isotopologues as the density tracers (e.g., \citealt{miotello18}; there is no prospect for spatially resolved HD measurements).  Some of the intrinsic degeneracies can be mitigated by modeling multiple species or line transitions simultaneously \citep{vanzadelhoff01,dartois03,schwarz16,zhang17,cleeves17}.  

Density measurements suffer the same ambiguities outlined for the masses (Section \ref{sec:mass}); namely, the uncertain conversion factors -- $\kappa_\nu$ for $\Sigma_{\rm s}$ and {\sf X}$_{\rm CO}$ (and isotopic fractions) for $\Sigma_{\rm g}$ -- and the potential for contamination by high optical depths.  There is also the added complexity that high $\Sigma_{\rm s}$ could produce an optically thick continuum that blocks the spectral line emission originating below (or behind) the continuum photosphere \citep[e.g.,][]{weaver18,dsharp9}.  If that effect is significant, robust estimates of $\Sigma_{\rm g}$ from line data will also require a simultaneous inference of $\Sigma_{\rm s}$ (a formidable challenge).  

Measurements of disk densities remain in an exploratory phase, with progress limited by data availability and quality, systematics in the methodology, and intrinsic degeneracies.  Some promising ideas for measuring $\Sigma_{\rm g}$ aim to get around the tracer abundance ambiguity, using line ratios that are directly sensitive to the volume density \citep[e.g.,][]{teague18c} or converting multiwavelength $R_{\rm mm}$ measurements and a simplified model for the aerodynamics of solids to an inference of the underlying density profile \citep{powell17}.

\subsection{Temperature} \label{sec:temp}

The thermal structure determines some fundamental reference scales, usually parameterized by the  sound speed, $c_{\rm s}$ ($\propto T^{0.5}$), and the pressure scale height, $H_p = c_s / \Omega_{\rm k} \propto (T r^3/M_\ast)^{0.5}$, where $\Omega_{\rm k}$ is the Keplerian angular velocity.  Moreover, it is intimately connected to the tracers of the disk material, since it controls the molecular excitation conditions, the vertical location of the scattering surface, and the spectral line and continuum intensities.

The temperature distribution depends on the irradiation of solids by the host star.  Small grains suspended in the disk atmosphere absorb starlight and then re-radiate some of that energy toward the midplane \citep{chiang97,dalessio98}.  That central, external energy deposition produces an increasing $T(z)$ \citep{calvet91} and a decreasing $T(r)$ \citep{kenyon87}.  Irradiation heating depends on the host star spectrum as well as the microphysical properties and vertical distribution of the solids \citep{dalessio99b,dalessio06,dullemond01,dullemond02}.  The latter is set by a balance between turbulent mixing and the solids-gas coupling \citep{dubrulle95}.  When the solids-to-gas ratio is low (at large $z$ or $r$), spectral line processes can super-heat the gas \citep{kamp04,bruderer13}.  A variety of secondary heating sources -- viscous dissipation \citep{dalessio98}, spiral shocks \citep{rafikov16}, radioactivity \citep{cleeves13}, external irradiation (e.g., from an envelope; \citealt{natta93,dalessio97}), or vertical structure perturbations (e.g., from self-shadowing; \citealt{dullemond04b}) -- can also contribute significantly to the temperature structure.  

The classical approach to constraining the temperature distribution is to forward-model the infrared SED.  Such modeling proposes a density and opacity distribution, simulates the propagation of energy through the disk, generates synthetic observables to compare with data, and iterates.  The fundamental challenges are the physical degeneracies in such modeling \citep{thamm94,heese17}; even if internally self-consistent, the models are not unique.  One way to mitigate some ambiguity is to fold additional (spatially resolved) data into the modeling circuit \citep[e.g.,][]{pinte08}.    
\begin{marginnote}[]
\entry{SED}{spectral energy distribution.}
\end{marginnote}

Another option relies on the spatial distribution of optically thick emission lines \citep{weaver18}.  With sufficient resolution, $T(r)$ can be measured in the vertical layer corresponding to the line photosphere \citep{rosenfeld13a,pinte18,dullemond19}.  Constraints on $T(r,z)$ are possible by probing intensities at different depths in the atmosphere using lines with a range of excitation conditions \citep{vanzadelhoff01,dartois03,schwarz16}.   That reconstruction effort can be supplemented with benchmarks in $T(r)$ from signposts of condensation fronts (snowlines), where volatiles are removed from the gas when they freeze onto grain surfaces (\citealt{qi11,qi19}).

\subsection{Dynamics} \label{sec:turb}

\begin{textbox}[t]\section{Observational Insights on Disk Magnetic Fields}
Magnetic fields are predicted to fundamentally alter the gas dynamics in disks, and thereby play important roles in shaping their structures and evolution.  But there are few concrete observational constraints available to inform magnetohydrodynamics models.  In principle, magnetic field morphologies can be measured from the linear polarization of mm continuum emission \citep{cho07,bertrang17} or molecular line emission \citep{goldreich81}.  So far,  efforts to measure the former have been frustrated by scattering \citep{kataoka15} and various alternative grain alignment mechanisms that can also polarize the continuum \citep[e.g.,][]{tazaki17,kataoka19}.  Linear polarization measurements of spectral lines from disks are expected soon.  The line-of-sight magnetic field strength (and topology) can potentially be measured with high resolution spectral line observations of circular polarization induced by Zeeman splitting (e.g., in CN hyperfine transitions; \citealt{brauer17a}).
\end{textbox}

Disks are profoundly affected by their fluid dynamics \citep{armitage11}.  The dominant factor in the kinematic structure of a disk is orbital motion, but important contributions are expected from magnetic fields (\citealt{turner14}; see the box above), viscous transport \citep{lyndenbell74}, pressure support \citep{weidenschilling77}, self-gravity \citep{rosenfeld13a}, and winds \citep{ercolano17}.  Random motions generated by turbulence are traditionally asserted as the source of a kinematic viscosity ($\nu_{\rm t}$) -- quantified with the coefficient $\alpha_{\rm t} = \nu_{\rm t} / c_{\rm s} H_p$ -- that controls accretion, mixing, and other diffusive processes.  Classical models of turbulence driven by the MRI \citep{balbus91} predict $\alpha_{\rm t} \approx 0.001$--0.01.  But a shifting theoretical paradigm now argues that the MRI is suppressed by non-ideal MHD effects over much of the disk \citep[e.g.,][]{bai13}, suggesting instead a system that is effectively laminar, $\alpha_{\rm t} < 0.001$.  
\begin{marginnote}[]
\entry{MRI}{magnetorotational instability.}
\entry{MHD}{magnetohydrodynamics.}
\end{marginnote}

Spatially and spectrally resolved observations of emission lines with a range of optical depths can be used to reconstruct the three-dimensional disk velocity field.  Typical observations are suitable for confirming that orbital motions dominate \citep{rosenfeld12b,czekala15,msimon17}, although measurements of non-Keplerian deviations are becoming available (Section \ref{sec:substructures}).  Constraints on turbulence are available from two approaches.  The first relies on a measurement of spectral broadening: an emission line profile has contributions from both thermal and non-thermal motions, with characteristic variances $2 k_{\rm B} T / m_j$ (where $k_{\rm B}$ is the Boltzmann constant and $m_j$ the mass of molecule $j$) and $\delta v_{\rm t}^2$, respectively.  With some knowledge of $T(r, z)$ (usually inferred jointly), resolved line measurements constrain $\delta v_{\rm t}$ in a given line photosphere layer.  Suitable data are only available in three cases.  In two of these (TW Hya and HD 163296), upper limits indicate sub-sonic turbulence ($\delta v_{\rm t} \lesssim 0.05 \, c_{\rm s}$) at $z \approx 1$--3\,$H_p$, corresponding to $\alpha_{\rm t} \lesssim 0.005$ \citep{hughes11,flaherty15,flaherty17,flaherty18,teague16,teague18b}.  A much broader $\delta v_{\rm t}$ ($\sim$0.$5 \, c_{\rm s}$) is found in the remaining case (DM Tau; \citealt{guilloteau12}).  Taken at face value, this implies vigorous turbulence ($\alpha_{\rm t} \gtrsim 0.1$) at a comparable altitude, or it may hint that the $T$ distribution is incorrect or other broadening mechanisms are at play. 

The second approach relies on the diffusive blurring of nominally ``sharp" features (Section \ref{sec:substructures}).  High resolution mm continuum observations offer {\it geometric} constraints on turbulent mixing, based on the height of the mm photosphere \citep{pinte16} or the radial widths of narrow ring features \citep{dsharp6}, that suggest $\alpha_{\rm t} \lesssim 10^{-3}$ near the midplane.  This methodology is complementary to the line broadening, with each approach probing different altitudes with orthogonal degeneracies (gas-particle coupling and the thermal structure of the gas, respectively).  Efforts to combine them can construct a more nuanced view of the spatial variation and origins of disk turbulence \citep[e.g.,][]{shi14}.

\section{DEMOGRAPHIC\ INSIGHTS} \label{sec:demographics}

The previous section highlighted the design and vetting of tools used to infer physical aspects of disk structures, as well as the intrinsic ambiguities and practical limitations that frustrate those inferences.  Those challenges are being confronted, with improved physical constraints following in step with the quality, volume, and diversity of the available data.  But assembling large, homogeneous catalogs of robust disk structure models is simply not practical.  Recognizing that, one imperative message from Section \ref{sec:structure} is that theoretical work in the physical domain ultimately needs to transform outputs into appropriate observational metrics: predictions and model tests should happen {\it in the data-space}.  

A proper demographic analysis requires a catalog of a given disk property (dependent variable) that is both large and spans a sufficient range in the external factors (independent variables) of interest.  The two empirical probes of disk structure properties that are simple enough to measure in large quantities today are the mm continuum luminosities ($L_{\rm mm}$) and sizes ($R_{\rm mm}$).  The remainder of this section synthesizes various data repositories to explore how these structure proxies depend on host masses (Section \ref{sec:mstar}), the local and global environments (Section \ref{sec:env}), and evolutionary diagnostics (Section \ref{sec:evol}).

\subsection{Links to Stellar Hosts} \label{sec:mstar}

Considerable attention in the field is devoted to probing connections between disk structures and their stellar hosts.  In particular, most theoretical work associated with star and planet formation presumes that fundamental physical principles like the conservation of mass and angular momentum could imprint some lasting relationships between the stellar host masses, $M_\ast$, and basic disk structure metrics like masses and sizes.  Some credence is lent to that emphasis from the clear $M_\ast$-dependencies that have been identified through demographic studies of the exoplanet population \citep[e.g., see][]{mulders18}.     

\begin{figure}[t!]
\includegraphics[width=\textwidth]{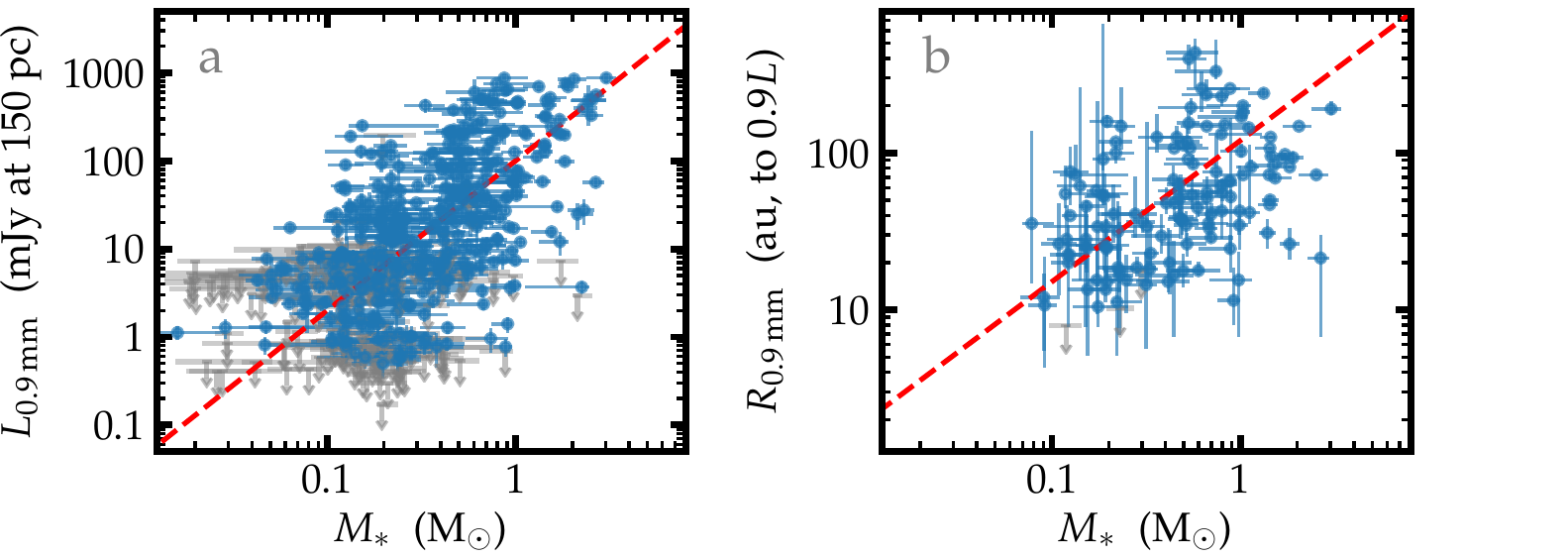}
\caption{(a) The correlation between $L_{\rm mm}$ and $M_\ast$ for disks (excluding multiple systems) in the Oph, Tau, Lup, Cha I, IC 348, and Upper Sco regions (see references and details on calculations in {\bf Figure \ref{fig:mass_dist}}).  Upper limits are shown as gray arrows.  The red line corresponds to the mean scaling relation inferred for the Oph, Tau, Lup, and Cha I disks only.  (b) The correlation between $R_{\rm mm}$ (defined as in {\bf Figure \ref{fig:sizes}}) and $M_\ast$, along with the mean scaling relation shown in red.}
\label{fig:mstar}
\end{figure}

Large mm continuum photometry catalogs for disks in a few nearby regions have sufficient dynamic range in $M_\ast$ to characterize any relationships with $L_{\rm mm}$  \citep{andrews13,ansdell16,pascucci16,barenfeld17}.  When excluding known multiple star systems (Section \ref{sec:env}), the regions with mean ages $\lesssim$\,3 Myr exhibit a consistent scaling relation, $L_{\rm mm} \propto M_\ast^{1.7\pm0.3}$ (for $M_\ast \ge 0.1$ M$_\odot$; the same scaling is found at $\lambda = 0.9$ or 1.3 mm), shown in {\bf Figure \ref{fig:mstar}a}.  The normalization indicates that a typical disk with a solar analogue ($M_\ast = 1$ M$_\odot$) host has flux densities of $\sim$100 or 40 mJy at 150 pc for $\lambda = 0.9$ or 1.3 mm, respectively.  There is considerable scatter around the mean $L_{\rm mm}$--$M_\ast$ relation, roughly a factor of three (0.5 dex) added dispersion in $L_{\rm mm}$ beyond the measurement uncertainties.  Some of that could be related to imprecise (or biased) $M_\ast$ estimates, though various physical origins are plausible.  There is a hint for Taurus disks \citep{ward-duong18,akeson19} that extending to $M_\ast < 0.1$ M$_\odot$ flattens the mean relation ($L_{\rm mm} \propto M_\ast^{1.2}$).  It is unclear if this is a real turnover or if it is unique to Taurus.

Assuming the emission is optically thin, this relation predicts a corresponding scaling between $M_{\rm s}$ and $M_\ast$ with a morphology that is sensitive to the behavior of the disk-averaged temperatures and opacities.  To date, all studies have presumed that $\langle \kappa_{\nu} \rangle$ is unrelated to $M_\ast$ (although without justification).  Various treatments of the $M_\ast$-dependence on $\langle T \rangle$ have been considered: \citet{andrews13} suggested that $\langle T \rangle \propto L_\ast^{1/4} \propto M_\ast^{1/2}$ based on simple irradiation heating arguments, while \citet{pascucci16} preferred the assumption that $\langle T \rangle$ is independent of $M_\ast$.  Given the measured behaviors of the size--luminosity (Section \ref{sec:sizes}; \citealt{tripathi17}) and size--$M_\ast$ ({\bf Figure \ref{fig:mstar}b}; \citealt{andrews18}) relations, simple irradiation heating should impose only a weak mass dependence on $\langle T \rangle$ (see \citealt{tazzari17}), in line with the \citeauthor{pascucci16} assumption and therefore predicting a steeper than linear $M_{\rm s}$--$M_\ast$ relation (i.e., a nearly $M_\ast$-independent link between $L_{\rm mm}$ and $M_{\rm s}$). 

The $R_{\rm mm}$--$M_\ast$ relation in {\bf Figure \ref{fig:mstar}b} is less pronounced, partly because the dynamic range in $M_\ast$ is limited (relative to the scatter) by resolution.  \citet{andrews18} estimated that a slightly sub-linear relationship was appropriate: the updated results here suggest $R_{\rm mm} \propto M_\ast^{0.9}$, consistent with a simple combination of the measured $L_{\rm mm}$--$M_\ast$ and $L_{\rm mm}$--$R_{\rm mm}$ scaling relations.  If the emission is optically thin, \citet{andrews18} demonstrated that such scaling behavior naturally follows if all disks have a similar mm optical depth profile (independent of $M_\ast$) with $\langle \tau_\nu \rangle \approx 0.4$, meaning $M_{\rm s}$ depends primarily on the disk size.  Alternatively, the same relationships would be produced if the emission is optically thick with an effective filling factor of $\sim$0.3, produced by spatially concentrating the high optical depths \citep{ricci12} or reducing the intensities by self-scattering from particles with high albedos \citep{zhu19}.  The scatter in these relations can be attributed to diversity in the underlying relation between the disk sizes and host masses, the mean optical depths, effective filling factors, or a combination of such effects.

\subsection{Environmental Effects} \label{sec:env}

\subsubsection{Dynamical Interactions}
Disk structures can be substantially shaped by dynamical interactions in their local environments.  The tidal perturbations that occur in multiple star systems are expected to be the most prevalent for the current catalog of disk observations (although see the box on the next page).  Multiplicity fractions are high, 30--50\%\ in the field \citep{raghavan10} and up to $\sim$70\%\ for the young clusters that inform most disk studies \citep[e.g.,][]{kraus11}.  Moreover, most stellar pairs have separations comparable to typical disk sizes ($\sim$10--100 au).  Simulations of the perturbations to disk structures induced by gravitational interactions in such systems find that individual disks in binaries are tidally truncated at $r \approx (0.2$-0.5$)\Delta$, where $\Delta$ is their mean separation  \citep{artymowicz94}; they generically predict that disks in close binaries are smaller, and therefore less massive, than their counterparts in wider binaries or around single stars.

\begin{figure}[t!]
\includegraphics[width=\textwidth]{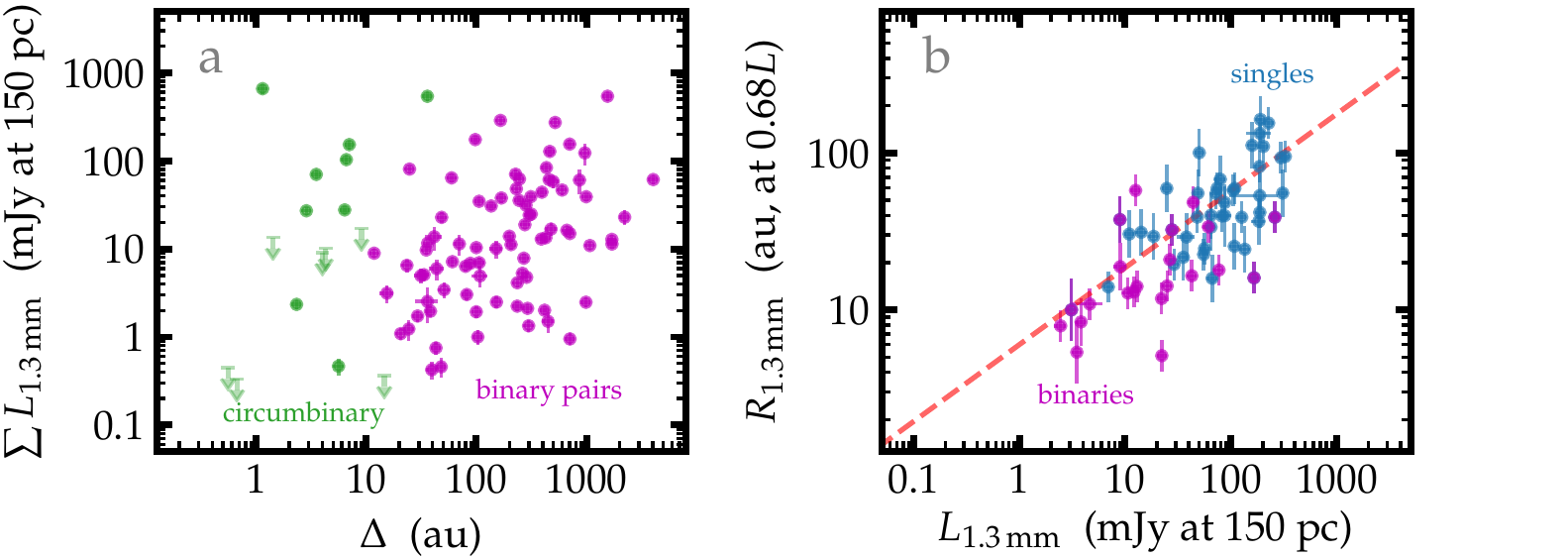}
\caption{(a) The summed continuum luminosities ($\lambda = 1.3$ mm) for binary pairs as a function of their projected separation.  Known circumbinary disks are marked in green.  The $L_{\rm mm}$-$\Delta$ behavior is qualitatively consistent with predictions of tidal truncation models.  (b) The $\lambda = 1.3$ mm $R_{\rm mm}$-$L_{\rm mm}$ relation (in this case, $R_{\rm mm}$ is defined as the radius encircling 68\%\ of $L_{\rm mm}$) for disks in binary systems (blue; \citealt{manara19}), compared with disks around similar, but single, hosts (orange; \citealt{dsharp1,long19}).  The red line marks the $\lambda = 0.9$ mm relation (Section \ref{sec:sizes}), renormalized with $L_{\rm mm} \propto \nu^{2.2}$ and $R_{\rm mm} \propto \nu^{0.3}$ (Section \ref{sec:solids}).}
\label{fig:binaries}
\end{figure}

There is some qualitative support for those predictions in the observations.  {\bf Figure \ref{fig:binaries}a} shows that the total $L_{\rm mm}$ in binary pairs marginally increases with their projected separation \citep{jensen94,harris12,akeson14}.  That behavior is convolved with the $L_{\rm mm}$-$M_\ast$ relation: \citet{akeson19} found the same shape for that relation applies for the individual components of binaries, but with a normalization offset.  The mean $L_{\rm mm}$ is 3--4$\times$ lower at the same $M_\ast$ for the binaries.  \citet{manara19} provided more support for the truncation hypothesis by comparing mm continuum emission sizes for disks in analogous subsamples.  {\bf Figure \ref{fig:binaries}b} demonstrates that $R_{\rm mm}$ for individual disks in binaries are $\sim$2$\times$ smaller than for a comparison set of disks around single stars.

However, \citet{manara19} found that the measured $R_{\rm mm}$ are too small compared with the truncation model predictions, given the projected separations (see also \citealt{harris12}).  The discrepancy could point to eccentric orbits or indicate that the models are inappropriate.  Those models presume co-planarity between the disks and stellar orbits, which is often not the case for the medium-separation binaries where $R_{\rm mm}$ estimates are tractable \citep{jensen14,williams14,tobin16,brinch16,alves19}.  Moreover, the models make predictions for the {\it gas} distribution, which is usually more extended than the solids (Section \ref{sec:sizes}).           

\begin{textbox}[t]\section{Unbound Dynamical Encounters}
Flyby encounters between unbound stars and their disks are a natural extension of the dynamical interactions experienced in binaries \citep{clarke93}.  The probability for such encounters is enhanced at early times, where the cluster environment has a higher local stellar density  \citep[e.g.,][]{bate18}.  \citet{pfalzner13} predict that $\sim$1 in 3 solar-type stars in an OB association experiences a close (100--1000 au) periastron passage within 1 Myr.  These flybys can substantially perturb disk structures, including the creation of spiral arms or tidal bridges \citep{cuello19}, truncation \citep{breslau14}, and warping or partial disruption \citep{xiang-gruess16}.  The key demographic prediction from these encounters is that {\it single} stars in clusters with higher stellar densities should host smaller, less massive disks \citep{dejuanovelar12,rosotti14}.  While a direct test of that hypothesis is not yet tractable (due to the current focus of ALMA surveys on nearby loose associations), there are signs of potentially related morphological features in individual systems, including possible tidal extensions \citep{winter18b} and spiral perturbations for disks in widely-separated binaries \citep{mayama12,rodriguez18,dsharp4}.
\end{textbox}

\subsubsection{External Photoevaporation}
Dynamical encounters are not the only environmental factors that alter disk structures.  The intense radiation produced by massive stars can heat the outer regions of nearby disks until the sound speed exceeds the escape velocity, generating considerable mass loss in a wind \citep{hollenbach94,alexander14}.  That externally-driven photoevaporative mass loss is validated with observations of ionization fronts associated with disks in the Orion Trapezium region \citep{johnstone98,storzer99}.  From a demographic perspective, photoevaporation models predict that disk sizes and masses should be lower in close proximity to massive stars.  Indeed, the mean $L_{\rm mm}$ drops within $\sim$0.03 pc of the massive star $\theta^1$ Ori C \citep{mann09,mann10,mann14}, corresponding to the region where ionization from its Lyman continuum radiation dominates \citep{johnstone98}.  \citet{eisner18} found that $L_{\rm mm}$ and continuum sizes increase with distance from $\theta^1$ Ori C, but are generally lower than for the disks in clusters without massive stars.  That behavior is consistent with models that suggest a larger region of influence on disk structures from less energetic (far-ultraviolet) radiation fields \citep{facchini16,ansdell17,vanterwisga19}.

\subsection{Evolutionary Signatures} \label{sec:evol}

\begin{figure}[t!]
\includegraphics[width=\textwidth]{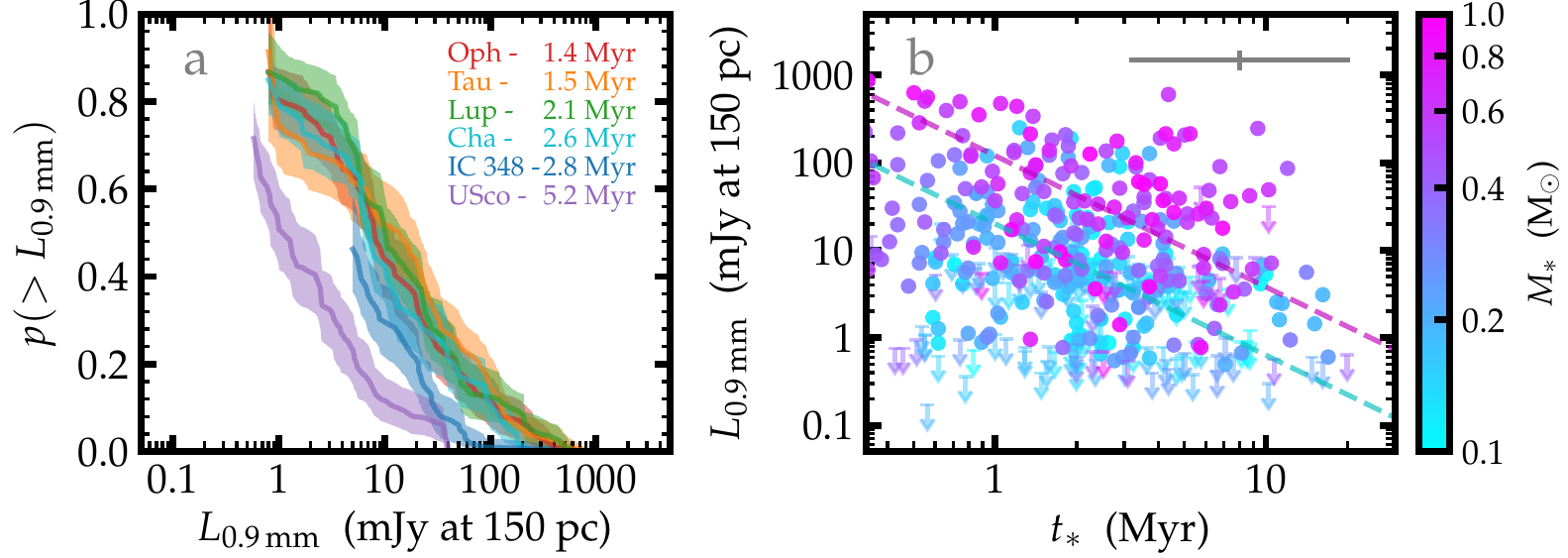}
\caption{(a) Comparisons of $L_{\rm mm}$ distributions in star-forming regions with different mean ages.  This is done in a Monte Carlo approach to control for $M_\ast$: 500 distributions are drawn for each region, with each a collection of 100 measurements chosen such that the host masses follow the same mass function (known multiple systems are excluded).  The colored bands show the 68\%\ confidence intervals of those draws.  There is a clear progression toward lower $L_{\rm mm}$ with mean cluster age (estimated from the parent samples).  (b) A more direct, individualized examination of the $L_{\rm mm}$--$t_\ast$ relation, that better highlights the challenges of such comparisons due to the intrinsic scatter.  Individual uncertainties are suppressed for clarity, but a mean error bar is shown in the top right.  The color scale tracks the $M_\ast$-dependence; the dashed line shows a $L_{\rm mm} \propto t_\ast^{-1.5}$ scaling (note, those scalings are {\it not} fits to the data, just a rough estimate to guide the eye).}
\label{fig:age}
\end{figure}

Much of the work on disk demographics focuses on the variations as a function of some metric of the elapsed time or evolutionary state of the system.  One option is direct, considering how disk properties depend on their stellar host ages, $t_\ast$.  While that seems natural, it is not trivial in practice because the ages are both imprecise and potentially inaccurate, due to biases in both the measurements and the models \citep[e.g.,][]{bell13}.  With those caveats in mind, common practice is to compare the distributions of a given disk probe in young star clusters with a progression of mean ages.  When controlling for $M_\ast$ and multiplicity, {\bf Figure \ref{fig:age}a} shows that the $L_{\rm mm}$ distribution shifts downward on $\sim$5 Myr timescales \citep{barenfeld17,ruiz-rodriguez18}.  A crude estimate of the decline suggests $L_{\rm mm} \propto t_\ast^{-1.5}$.  There is evidence that the shape of the $L_{\rm mm}$ distribution changes, manifested as a steepening in the $L_{\rm mm}$--$M_\ast$ relationship over time \citep{pascucci16,barenfeld17}.  {\bf Figure \ref{fig:age}b} shows the same information in individual datapoints.         

Some, perhaps all, of this evolution in $L_{\rm mm}$ is associated with changes in the continuum size--luminosity relationship: $R_{\rm mm}$ is generally smaller for disks in older clusters (\citealt{barenfeld17_sizes,hendler19}).  Such behavior indicates that the growth and migration of disk solids are key factors driving these demographic trends, rather than wholesale $M_{\rm s}$ changes (\citealt{tripathi17,rosotti19b}; Section \ref{sec:solids}).  But analyses like these tell only part of the story.  Focusing solely on systems that show excess infrared emission introduces a form of survivor bias by not accounting for the fact that the disk (infrared excess) fraction also decreases with $t_\ast$ \citep{haisch01,hernandez07}.  With that in mind, the combined effects of evolution are clearly under-estimated.  

An alternative approach can help mitigate that survivor bias.  The idea is to track how disks change as a function of their SED shape, an empirical diagnostic of the evolutionary state of the circumstellar material.  Generally, this sequence where the SED peak moves progressively to shorter $\lambda$ reflects the dissipation of the envelope (Class 0 $\rightarrow$ I $\rightarrow$ II) and disk (Class II $\rightarrow$ III; e.g., \citealt{williams11}).  The $L_{\rm mm}$ (or $M_{\rm s}$) distributions again shift downward along this evolutionary sequence \citep{aw07b,sheehan17,tychoniec18,williams19}.  It is not easy to measure disk properties in the embedded phases (Class 0/I), due to both the technical challenge of disentangling emission from the envelope \citep{tobin15} and the potential for younger disks to be intrinsically small \citep{segura-cox18,maury19}.  Most estimates find little evolution in $L_{\rm mm}$ during the Class 0 to I transition, even with a considerable decrease in envelope mass \citep{jorgensen09,segura-cox18,andersen19}.  

Despite some of the benefits of this latter evolutionary axis, it is difficult to contextualize the results without reference to a quantitative timeline.  Moreover, there is potential to make unfair comparisons that are not able to control for orthogonal relationships (e.g., a $M_\ast$ dependence, since stellar properties for Class 0/I sources are unknown) or sample completeness (e.g., large mm Class III surveys are unavailable).

\section{THE\ EVOLUTION\ OF\ DISK\ SOLIDS} \label{sec:solids}

The physical origins of the demographic behaviors that were outlined in the previous section are presumably closely related to the growth and migration of the disk solids.  However, the vast scope and complexity of that evolution is daunting.  To generate a population of planetesimals suitable for assembling a planetary system, the sub-$\mu$m dust grains incorporated into the disk at its formation epoch need to grow $>$\,12 orders of magnitude in size within a few Myr.  This section highlights the basic theoretical framework developed to understand these processes (Section \ref{sec:growth-theory}), explores the observational constraints (Sections \ref{sec:opacities} and \ref{sec:dustevol}), and considers the implications of some persistent obstacles (Section \ref{sec:planetesimals}).

\subsection{Standard Theoretical Picture} \label{sec:growth-theory}

The two key elements required to model the evolution of disk solids are prescriptions for their coupling to the fluid motions of the gas \citep{nakagawa86} and the outcomes of particle collisions \citep{guttler10}.  Standard models start with small dust grains distributed homogeneously within a {\it smooth} gas disk, where the pressure ($P$) decreases monotonically with $r$ and $z$.  Turbulent diffusion is described with a simple viscosity prescription for fixed $\alpha_{\rm t}$.  The small dust is well coupled to the gas, and so acquires low relative velocities through diffusive motions that result in gentle collisions that promote growth to porous aggregates \citep{henning96,dominik97}.  Those aggregates settle toward the midplane \citep{dubrulle95}, where the growth sequence continues.  The material properties (internal structure, charge state, ice coating), sizes, and relative velocities of the impactor and target solids determine whether a collision is productive (mass transfer; \citealt{teiser09}), neutral (bouncing; \citealt{zsom10}), or destructive (fragmentation, erosion; \citealt{birnstiel10,krijt15}).  Simulations indicate that growth continues until collisions become destructive \citep{dullemond05} or the local particle population is depleted by radial migration \citep{takeuchi02,brauer07,birnstiel09}.  For typical disk parameters, the latter effect dominates.                  

For a smooth disk, pressure support generates an additional outward force on a parcel of gas that effectively slows its orbital motion \citep{whipple72}.  The radial migration (``drift") of solids occurs once particles reach a size where they start to aerodynamically decouple from the gas; once disconnected from the pressure support of the gas, the particles spiral inwards toward the global maximum in $P$ \citep{weidenschilling77}.  The timescales for that migration are much shorter than the collision timescales, thereby inhibiting further growth at that location \citep{takeuchi05,brauer08}.  As a guide, drift is typically most efficient for pebbles (mm/cm sizes) at $r \approx 10$--100 au.  

The combined effects of growth and migration -- both vertically (settling) and radially (drift) -- profoundly influence the properties of disk solids \citep{testi14}.  The simplest distillation of the key predictions in this standard framework is that disks should exhibit pronounced, negative spatial gradients (i.e., decreasing with $r$ and $z$) in their mean particle sizes and solids-to-gas mass ratios, such that larger solids at higher concentrations (relative to the gas) are found near the midplane \citep{dullemond04a,dalessio06} and closer to the host star \citep{birnstiel09,birnstiel15,birnstiel14}.

\begin{textbox}[t]\section{High-Dimensional Complexity in Particle Properties}
The limited scope of the metrics explored in Section \ref{sec:opacities} reflect the over-simplified emphasis in the literature.  The reality is that many other factors can influence the absorption and scattering properties of the particles, and therefore the key observables \citep[e.g.,][]{min16}.  This high-dimensional complexity includes mineralogical compositions \citep{henning96,cuzzi14,woitke16}, asphericity \citep{bertrang17b}, temperature-dependent refractive indices \citep{boudet05}, the methodology for mixing dielectric properties in composite particles \citep{dsharp5}, and more sophisticated particle size distributions \citep{birnstiel11}, to name only a few.  While these issues could change the details, the qualitative behaviors should be generally preserved.  However, when confronted with subtle discrepancies or tensions (e.g., Section \ref{sec:tension}), a wider exploration of these other factors should be seen as a priority.  
\end{textbox}

\subsection{Metrics of Particle Properties} \label{sec:opacities}

In principle, those key predictions can be measured observationally.  But designing the appropriate experiments and then interpreting the measurements requires a nuanced understanding of how particle properties are translated into disk tracers.  The interactions of solid particles with radiation depend on their bulk properties, including compositions \citep{pollack94}, morphologies \citep{henning96}, and especially sizes \citep{miyake93} -- but see also the box above.  Those properties are encoded in the (absorption) opacities ($\kappa_\nu$), albedos ($\omega_\nu$), and polarizations ($\mathcal{P}_\nu$) of the particle ensemble.\footnote{The phase angle variations of $\omega_\nu$ and $\mathcal{P}_\nu$ also contain information about the particles.}  Measurements of the thermal continuum and scattered light reflect the convolution of the physical conditions of the solids and the behaviors of \{$\kappa_\nu$, $\omega_\nu$, $\mathcal{P}_\nu$\}.

The morphology and size distributions for a population of solids have the most significant effects on the observables.  Technically, morphologies are affected by both shape and internal structure (porosity), but the former is often ignored.  The porosity is parameterized by a volume filling factor $f_{\rm s}$ ($=$1 for compact particles).  Particle size distributions are usually approximated as power-laws, $n(a) \propto a^{-q}$ for sizes (particle radii) $a \in [a_{\rm min}, a_{\rm max}]$, with indices comparable to expectations for a collisional cascade ($q \approx 3.5$; \citealt{dohnanyi69}) or a more top-heavy variant ($q \approx 2.5$; e.g., \citealt{birnstiel11}).  

\begin{figure}[t]
\includegraphics[width=\textwidth]{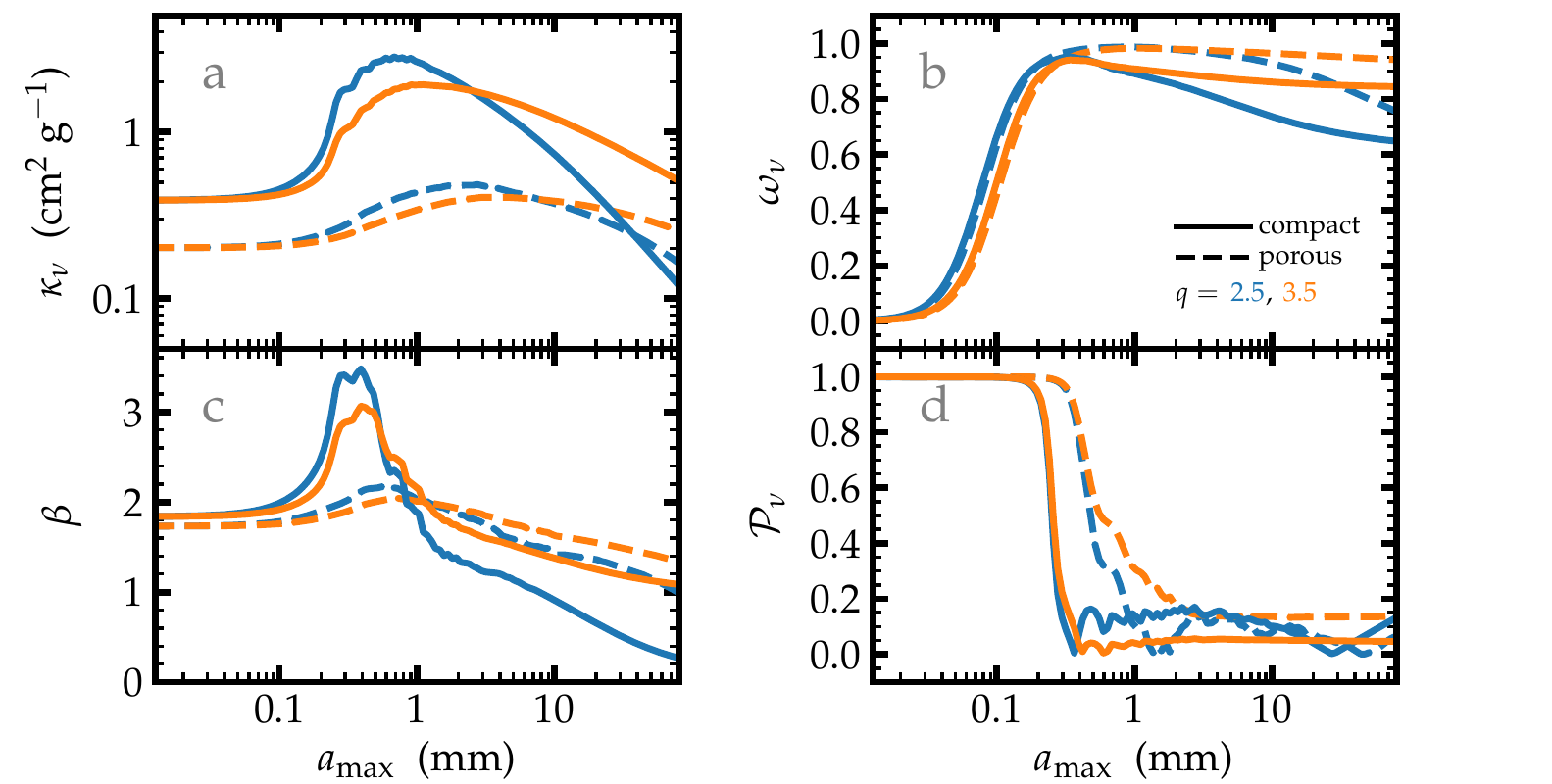}
\caption{The variations of the $\lambda = 1.3$ mm (a) absorption opacity, (b) albedo (accounting for the mean scattering angle), (c) absorption opacity spectral index (between 1.3 and 3 mm), and (d) polarization fraction as a function of the maximum particle size.  Each panel contains four curves, showing different power-law size distributions ($q = 2.5$ in blue, $q = 3.5$ in orange) and porosities (compact grains with $f_{\rm s} = 1$ as solid, and $f_{\rm s} = 0.5$ as dashed).  These behaviors were calculated with a standard Mie scattering code for the assumptions outlined by \citet{dsharp5}.}   
\label{fig:opac}
\end{figure}

Because of the (presumed) low optical depths, much of the work on particle properties in disks is conducted at mm/cm wavelengths.  There, $a_{\rm min}$ is irrelevant and the opacity spectrum can be approximated as a power-law, $\kappa_\nu \propto \nu^\beta$.  {\bf Figure \ref{fig:opac}} illustrates how \{$\kappa_\nu$, $\beta$, $\omega_\nu$, $\mathcal{P}_\nu$\} respond to the particle properties \{$a_{\rm max}$, $q$, $f_{\rm s}$\} at $\lambda = 1.3$ mm for the assumptions of \citet{dsharp5}.  The behavior at other wavelengths is qualitatively similar, with the main features shifted for an $a_{\rm max} \sim \lambda / 2\pi$ scaling.  When $a_{\rm max} \ll \lambda$, $\kappa_\nu$ is independent of size, $\beta$ is high ($\sim$1.7, as for the small dust grains in the ISM; \citealt{finkbeiner99}), and scattering is negligible ($\omega_\nu \approx 0$, though $\mathcal{P}_\nu$ is high).  When $a_{\rm max} \gg \lambda$, $\kappa_\nu$ decreases with $a_{\rm max}$ at a rate that depends on $q$ (lower $q$ means a steeper fall-off; e.g., \citealt{ricci10a}), $\beta$ is lower (scaling roughly with $q$; \citealt{draine06}), albedos are high (larger $q$ implies higher $\omega_\nu$), and $\mathcal{P_\nu}$ is low.  When $a_{\rm max} \sim \lambda$, resonances drive up $\kappa_\nu$, $\beta$, and $\omega_\nu$, while $\mathcal{P}_\nu$ drops precipitously.  Porosity dampens the resonant amplifications in $\kappa_\nu$ and $\beta$, but can enhance $\omega_\nu$ and $\mathcal{P}_\nu$, and generally modifies the $a_{\rm max} \gg \lambda$ behavior \citep{kataoka14}.  

It is worthwhile to specifically address the apocryphal notion that (optically thin) mm continuum emission traces particles with $a \sim \lambda$.  A more accurate statement is that the emission is most {\it efficient} in that case, since this corresponds to the resonant peak in $\kappa_\nu$ and therefore gives the most emission per mass.  However, all sizes still contribute, and that creates an intrinsic ambiguity: $\kappa_\nu$ can be arbitrarily low if larger solids are present.  An observational constraint on $\beta$ only sets a lower bound on $a_{\rm max}$, since $\beta$ effectively saturates once $a_{\rm max} \gg \lambda$.  That, in turn, sets an upper bound on $\kappa_\nu$, and correspondingly a lower bound on the mass-related quantities (namely, $\Sigma_{\rm s}$ or $M_{\rm s}$).

\subsection{Measurements of Particle Growth and Migration} \label{sec:dustevol}

\subsubsection{Scattered Light and the Infrared SED} \label{sec:growth-IR}
Optical and near-infrared images demonstrate that the starlight reflected from disk surfaces is typically faint (low $\omega_\nu$; \citealt{fukagawa10}), gray or red \citep{weinberger02,schneider03}, and forward-scattered (\citealt{quanz11,mulders12}).  Taken together, those properties indicate dust aggregates with $a_{\rm max} \gtrsim 10$ $\mu$m in disk atmospheres, representing the early steps in the growth sequence or possibly tracing collision fragments mixed up from the midplane.  Similar conclusions are drawn from the shapes of solid-state emission features in the mid-infrared, though isolating the inner disk with an interferometer is essential for robustly assessing the more processed grains located in the inner disk \citep{vanboekel04}.  

Direct measurements that trace the settling of dust aggregates toward the disk midplane are difficult due to the small intrinsic extent of the vertical dimension (with characteristic aspect ratios $z/r \lesssim 0.1$).  Very high resolution mm continuum observations of edge-on disks are expected to provide decisive constraints on settling in the near future \citep[e.g.,][]{boehler13,louvet18}.  For now the effects are identifiable with indirect probes, like the morphology of the infrared SED.  Settling depletes particle densities relative to the gas in the disk atmosphere, reducing the infrared opacity and associated continuum emission below expectations from models that assume gas and dust are well-mixed  \citep{dullemond04b,dalessio06}.  The suppression of the infrared SED inferred from those models suggests that the dust-to-gas ratio is depleted 10--100$\times$ in disk atmospheres \citep[e.g.,][]{furlan11}.  Analogous evidence can be retrieved from multiwavelength scattered light images: settling induces a vertical stratification of particle sizes ($a_{\rm max}(z)$ is decreasing), and the corresponding gradient in $\omega_\nu$ makes the height of the scattering surface decrease with $\lambda$ \citep{pinte07,duchene10,mccabe11,muro-arena18}.

\subsubsection{Millimeter Continuum Spectrum} \label{sec:growth-mm}
The mm continuum emission offers the most discriminating probes of particle properties near the disk midplane.  In the optically thin limit, the intensity scales like $I_\nu \propto \kappa_\nu \, B_\nu(T) \, \Sigma_{\rm s}$.    But since $\kappa_\nu$ cannot be determined uniquely, information about the particle properties is only accessible through the shape of the spectrum, quantified by the spectral index $\varepsilon$ (where $I_\nu \propto \nu^{\varepsilon}$), with $\varepsilon \approx \varepsilon_{\rm Pl} + \beta$ a sum of contributions from the Planck function ($B_\nu \propto \nu^{\varepsilon_{\rm Pl}}$, where $\varepsilon_{\rm Pl} \approx 1.7$--2.0 for $T > 15$ K) and the opacity spectrum ($\kappa_\nu \propto \nu^\beta$).  Resolved measurements of $\varepsilon(r, z)$ can test the predicted spatial segregation of particle sizes.  Larger particles have smaller $\beta$ ({\bf Figure \ref{fig:opac}}), and therefore smaller $\varepsilon$: the hypothesis is that $\varepsilon$ {\it increases} (the spectrum steepens) with $r$ and $z$.         

\begin{figure}[t!]
\includegraphics[width=\textwidth]{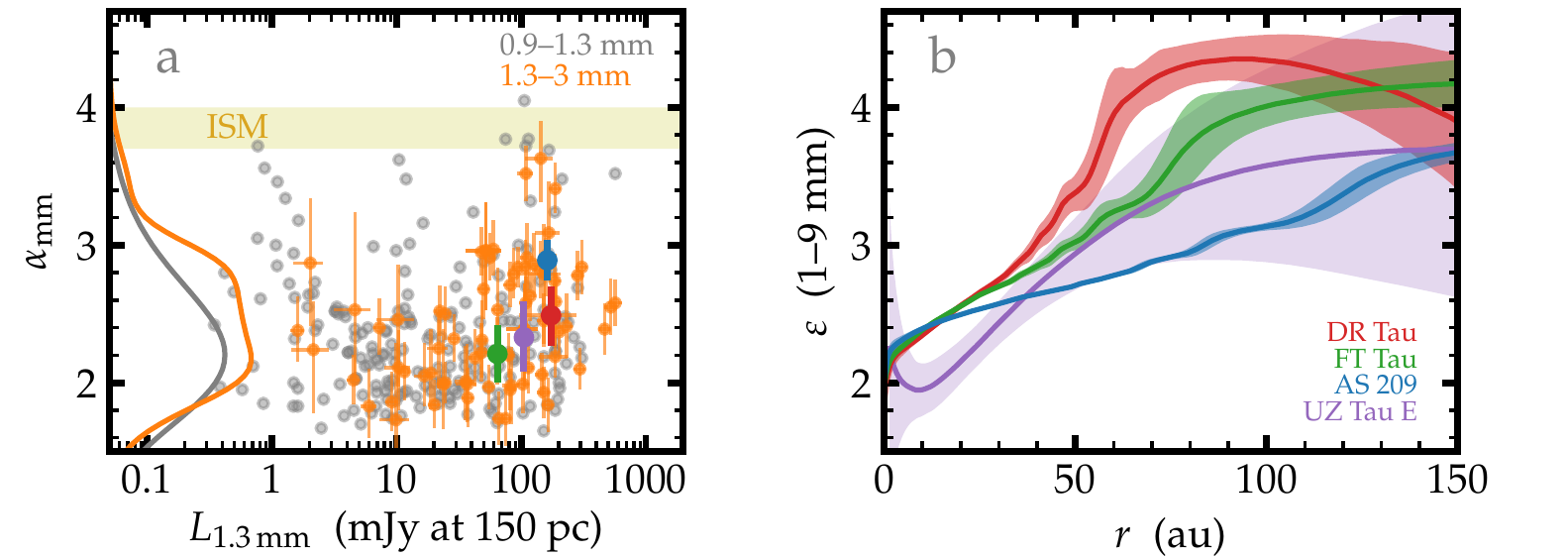}
\caption{(a) The disk-integrated spectral indices, $\alpha_{\rm mm}$, as a function of $L_{\rm mm}$ at 1.3 mm.  Data were collected from the photometry surveys mentioned in the {\bf Figure \ref{fig:mass_dist}} caption and supplemented with additional data when available \citep{ricci10b,ricci10a,ricci12,ricci13,ricci14,lommen10,harris12,cieza15,testi16,vanderplas16,cox17,pinilla17b,ward-duong18}.  Most spectra are much shallower than in the ISM, indicating particle growth and/or high optical depths.  The $\alpha_{\rm mm}$ distributions (accounting for uncertainties) between 1.3--3 mm (orange) and 0.9--1.3 mm (gray, with $\sim$4$\times$ as many measurements, but less useful frequency leverage; uncertainties are suppressed for clarity, but are typically $\sim$0.4 in $\alpha_{\rm mm}$) are shown along the ordinate axis.  (b) The radial variations in the continuum spectral indices, between $\lambda = 1$ and 9 mm, inferred from modeling resolved multiwavelength continuum observations of four individual disks \citep{tazzari16,tripathi18}.  Shaded areas show 68\%\ confidence intervals around the posterior means marked with darker curves.  These targets are marked with the corresponding colors in panel (a).}
\label{fig:mm_indices}
\end{figure}

In practice, the disk-integrated spectral index $\alpha_{\rm mm}$ (where the flux density $F_\nu \propto \nu^{\alpha_{\rm mm}}$) is a much more common metric in the literature.  As shown in {\bf Figure \ref{fig:mm_indices}a}, measurements find that $\alpha_{\rm mm} \approx 2$--3 in the $\lambda \approx 1$--3 mm range (the mean $\alpha_{\rm mm} = 2.2\pm0.3$ for $\lambda = 0.9$--1.3 mm), with a modest preference for steeper spectra at larger $L_{\rm mm}$ \citep{beckwith91,aw05,aw07a,ricci10a,ricci10b,ansdell18}.  There are tentative hints that $\alpha_{\rm mm}$ also increases with $R_{\rm mm}$ and $M_\ast$ (as might be expected; see Sections \ref{sec:sizes} and \ref{sec:mstar}), but the scatter is large and selection effects may dominate.  Taken at face value, the measured $\alpha_{\rm mm}$ indicate shallow opacity spectra ($\beta \lesssim 1$), and therefore large particles (e.g., $a_{\rm max} \gtrsim 10$ cm for the models in {\bf Figure \ref{fig:opac}}).  

However, unresolved spectral index measurements gloss over some important complexities.  Note that $\alpha_{\rm mm}$ is not a disk-averaged $\varepsilon$ ($\alpha_{\rm mm} \neq \langle \varepsilon \rangle$).  The interpretation of $\alpha_{\rm mm}$ in the context of spatial variations in the continuum spectrum is ambiguous.  Constraints on $\varepsilon(z)$ are difficult, again due to the intrinsically low aspect ratios of disks.  That said, the rare limits on the vertical extent of the mm continuum are qualitatively consistent with the particle size segregation predicted by settling models with low $\alpha_{\rm t}$ \citep{guilloteau16,pinte16}.  Measurements of $\varepsilon(r)$ that trace the combination of growth and radial drift are more practical.  The common approach is to reconstruct $\varepsilon(r)$ from ratios of model fits to the multiwavelength $I_\nu(r)$ profiles \citep{isella10,guilloteau11}.  {\bf Figure \ref{fig:mm_indices}b} shows some examples.  A condensed alternative considers the wavelength dependence of the continuum sizes: an increasing $\varepsilon(r)$ implies a decreasing $R_{\rm mm}(\lambda)$ \citep{tripathi18}.  In either case, such analyses infer that $\varepsilon(r)$ increases from $\sim$2 in the inner disk (tens of au) to $\ge$\,3 at larger $r$, corresponding to $\beta$ growing from $\sim$0 to $\ge$\,1 and $a_{\rm max}$ decreasing from $\gtrsim$\,cm to $\lesssim$\,sub-mm sizes \citep{perez12,lperez15,menu14,tazzari16}.

\subsubsection{Polarization}
Some complementary constraints on particle sizes are available from the linear polarization of self-scattered mm continuum emission \citep{hughes09b,kataoka15}.  As illustrated in {\bf Figure \ref{fig:opac}}, the albedo and polarization change precipitously in opposite directions near $a_{\rm max} \approx \lambda / 2\pi$.  The narrow shape of the product $\omega_\nu \mathcal{P}_\nu$ implies that polarization from scattering is only produced for a limited range of particle sizes.  Measurements of the wavelength of peak $\mathcal{P}_\nu$ set a stringent limit on $a_{\rm max}$.  Resolved observations of polarized emission at $\lambda = 0.9$--1.3 mm find that the bulk morphologies of the polarization vectors are consistent with model predictions for scattering \citep{kataoka16a,yang17,stephens17,hull18,bacciotti18,dent19}.  For a few disks, multiwavelength $\mathcal{P}_\nu$ measurements indicate $a_{\rm max} \approx 0.1$ mm \citep{kataoka16b,ohashi18}, considerably lower than inferred from the spectral indices.

\subsubsection{Tension and the Optical Depth Caveat} \label{sec:tension}
The explanation for this apparent discrepancy in the characteristic $a_{\rm max}$ inferred from the spectral indices and polarization properties of the mm continuum emission is not yet clear.  One potential reconciliation is that the comparison itself could be misleading due to spatial variations in one or both of the tracers.  For example, the outer regions of disks tend to have $\varepsilon \gtrsim 3$, which could be consistent with the polarization-based size constraints if much of the $\mathcal{P}_\nu$ behavior is produced at larger radii.  But perhaps a simpler and more compelling possibility is that the assumption of low optical depths used to simplify the interpretation of the mm continuum emission is invalid.  High optical depths suppress the continuum spectral index, with $\varepsilon \approx 1.5$--2.5, depending on the local temperature and the spectral variation of the albedo if scattering is important \citep{zhu19,liu19}.  There is still information about the particle sizes (in the $\tau_\nu \approx 1$ photosphere layer) available from $\varepsilon$ in this case, but the quantitative limits on $a_{\rm max}$ could indeed be very different than would be inferred in the optically thin limit.  Though this possibility had previously been considered \citep{ricci12}, it is worth revisiting in the context of more detailed measurements of the disk emission (see Section \ref{sec:substructures}).

\subsubsection{Comparisons with Spectral Line Emission}
Observational tests of the prediction that the solids-to-gas ratio decreases with $r$ are more challenging.  Quantitative measurements of that ratio are impractical, given the ambiguities associated with measuring $\Sigma_{\rm s}$ and $\Sigma_{\rm g}$ (Section \ref{sec:sigma}).  Instead, investigations rely on a qualitative approach analogous to the SED constraints on the vertical variation of the dust-to-gas ratio discussed in Section \ref{sec:growth-IR}.  The strategy is to negate the hypothesis of a radially constant solids-to-gas ratio by demonstrating that such models cannot simultaneously explain the intensity profiles of both the mm continuum and a bright spectral line \citep{isella07,panic09,andrews12}.  The argument is that the size discrepancy between the continuum and line emission ({\bf Figure \ref{fig:sizes}b}) is an indirect indicator that $\Sigma_{\rm s}$ and $\Sigma_{\rm g}$ have different shapes.          

Realistically, such a comparative analysis is not robust enough to be quantitative.  There are legitimate concerns about radiative transfer effects, since the tracers being compared have very different optical depths \citep{hughes08,trapman19}.  Moreover, it is not easy to disentangle the signatures of a solids-to-gas ratio that decreases with $r$ from the accompanying changes in $\kappa_\nu(r)$ \citep{facchini17,rosotti19a}.  \citet{trapman19} argued that $R_{\rm CO} / R_{\rm mm} \gtrsim 4$ is an unambiguous indicator of growth and drift for smooth disks.  The typically lower values of that ratio could still be consistent with that scenario (detailed modeling would be necessary), but might also reflect deviations from a smooth gas disk, where $R_{\rm mm}$ is effectively increased by slowed particle migration at local $P$ maxima while $R_{\rm CO}$ is unaffected (see Section \ref{sec:substructures}).

\subsection{Toward Planetesimals} \label{sec:planetesimals}

The measurements outlined above are in good {\it qualitative} agreement with the standard theoretical predictions for the evolution of disk solids.  This empirical support suggests that the basic physical ingredients in the models are appropriate.  However, there are two important {\it quantitative} problems with the framework.  The first is subtle: the predicted spatial segregation of particle sizes is generically more extreme than is implied by measurements of $\varepsilon(r)$ \citep[e.g.,][]{tripathi18}.  The same problem appears as an over-prediction of the $\alpha_{\rm mm}$ distribution with respect to observations \citep{birnstiel10,pinilla13}, and a difficulty in reproducing the high end of the $R_{\rm mm}$--$L_{\rm mm}$ correlation \citep{tripathi17,rosotti19b}.  Put simply, the predicted evolution is too fast to account for the data.  The second problem is related, but more striking: the models do not produce planetesimals, or even $\gtrsim$\,meter-sized bodies, within the timeframe associated with disk dispersal ($\sim$5--10 Myr).  Inside a few au, this latter issue is perhaps associated with incomplete physics in the models \citep{laibe12,okuzumi12,windmark12b,windmark12a}.  

Ultimately, both of these problems are consequences of the predicted (high) efficiency for the radial migration of solids.  The next section reconsiders an elegant solution to both problems, achieved by relaxing the standard assumption of a smooth gas disk.

\section{SUBSTRUCTURES} \label{sec:substructures}

Until recently, nearly all of the constraints on disk structures were derived from observations with relatively coarse spatial resolution, $\gtrsim$\,15--20 au.  As is frequently the case, improved facilities and instrumentation have precipitated a dramatic shift in the field, with a new emphasis on the prevalence of fine-scale features, or {\it substructures}, in these disks.  Despite the narrowed focus on these details, important new insights have emerged that are re-shaping how disk properties are interpreted and contextualized more generally.

\subsection{Resolving the Drift Dilemma} \label{sec:fix_drift}

Substructures can reconcile the two fundamental problems associated with the migration of solids in the classical theory (Section \ref{sec:planetesimals}).  To explain how, it helps to revisit the cause of the migration.  The force balance between gravity, rotation, and pressure support determines the orbital motion of the gas disk.  The contribution associated with pressure support is proportional to the gradient d$P$/d$r$.  The standard assumption of a {\it smooth}, monotonically decreasing $P(r)$ implies that d$P$/d$r$ is always negative, and therefore the gas orbits at sub-Keplerian velocities. For solids that decouple from the gas, drag extracts orbital energy and imparts a radial velocity directed inwards, toward the $P$ maximum at the inner disk edge.  The key problem is that this radial drift is too efficient \citep{takeuchi02,takeuchi05}.

However, if $P(r)$ is not monotonic but instead has {\it local} maxima, there are corresponding modulations to the gas dynamics with striking consequences.  Exterior to a local maximum, the standard physical scenario applies: d$P$/d$r < 0$, gas velocities are sub-Keplerian, and drifting particles move inwards.  But just interior to a maximum, d$P$/d$r > 0$ and the gas motion is super-Keplerian.  In that case, particles are instead pushed outwards.  Interactions with this perturbed gas flow drive particle migration to converge toward the local pressure maximum.  At the maximum there is no pressure gradient (d$P$/d$r = 0$ by definition), so the gas motion is Keplerian and the solids do not drift.  A sufficiently steep $P$ modulation with limited diffusion can effectively ``trap" solids by slowing or halting their migration.   

These substructures in the gas pressure distribution are essential ingredients for reconciling the drift and planetesimal formation timescale problems.  If distributed throughout the disk, perturbations to $P(r)$ can alleviate the drift timescale problem by stalling particle migration \citep{pinilla12a,pinilla13}.  Moreover, the resulting localized particle concentrations can attain solids-to-gas ratios that approach unity \citep[e.g.,][]{yang17b}, thereby creating favorable conditions for the rapid conversion of pebbles into planetesimals through the streaming instability \citep{youdin05,johansen07a} or direct gravitational collapse \citep{goldreich73,youdin02}.

\subsection{Potential Physical Origins} \label{sec:origins}

The hypothesis that substructures are both elemental disk characteristics and fundamental aspects of the planet formation process is agnostic about their physical origins.  However, a remarkable variety of ways to generate substructures that trap migrating solids (or otherwise perturb their migration) have been proposed in the literature.  The discussion below highlights these mechanisms, coarsely grouped into three general categories.  The schematics in {\bf Figure \ref{fig:sub_cartoon}} illustrate how some of these mechanisms are manifested as small-scale perturbations to the distributions of gas and solids in the disk.

\subsubsection{Fluid Mechanics} \label{sec:fluids}
\begin{figure}[t!]
\includegraphics[width=\textwidth]{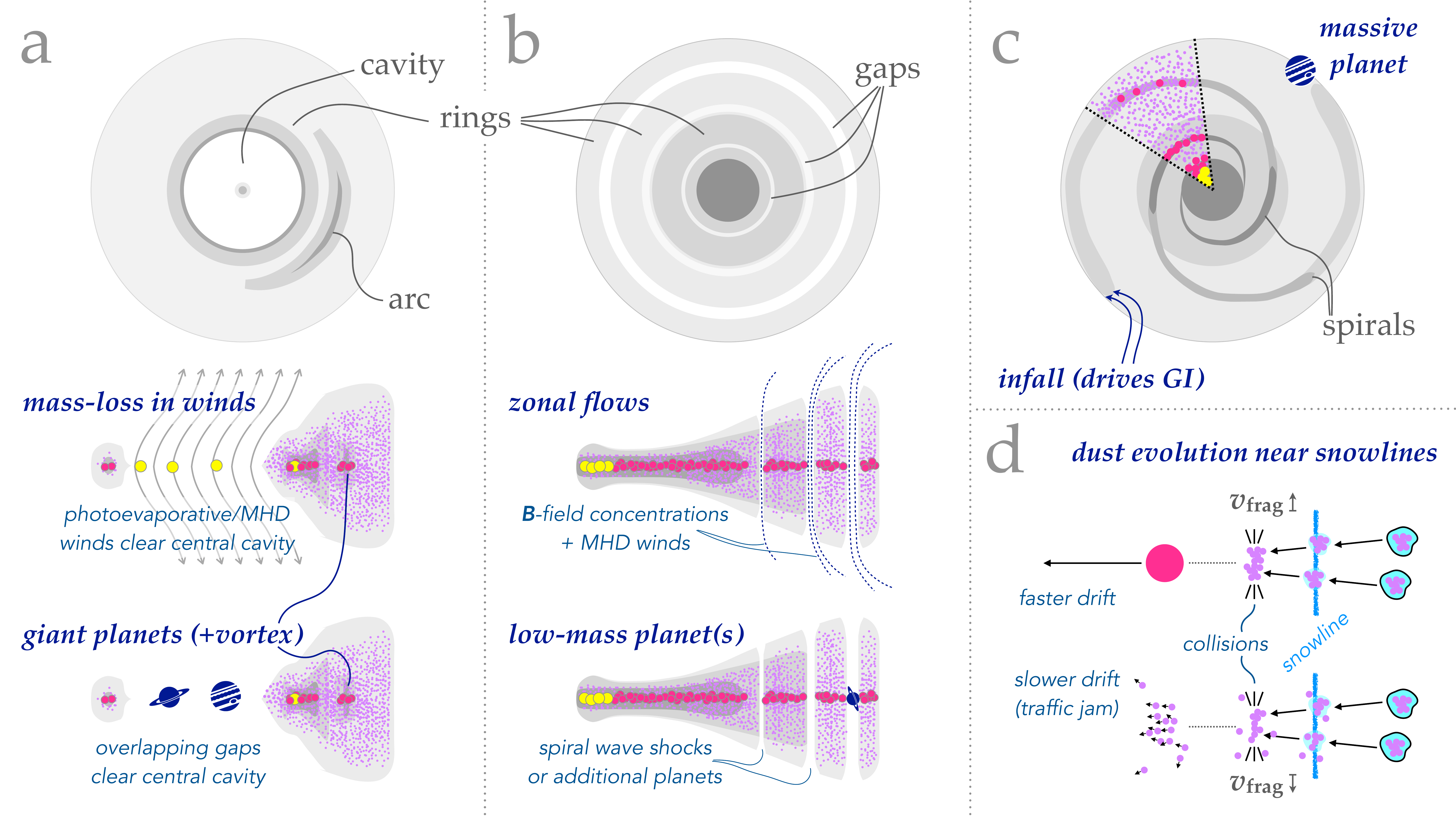}
\caption{Schematic illustrations of the substructures generated by various physical mechanisms.  As in {\bf Figure \ref{fig:cartoon_overview}}, grayscale denotes gas densities ($\propto P$) and representative solid densities are marked with exaggerated symbol sizes and colors.  (a) A schematic of a ring/cavity substructure morphology with a pronounced arc feature generated by a vortex.  The two side views represent the behavior for a disk with substantial mass-loss in a photoevaporative or MHD-driven wind (top) or a series of giant planets (bottom), both of which effectively diminish $\Sigma_{\rm g}$ in a central cavity.  The sharp density contrast at the cavity edge can trap particles in a ring and potentially generate a vortex.  (b) A schematic of the ring/gap substructure morphology, with similar behavior produced by the magnetic field concentrations inherent in MHD zonal flows (top) and the perturbations from interactions between lower mass planets and a relatively inviscid disk (bottom).  (c) A simplified diagram of the spiral wave perturbations that could be produced by the global GI driven by remnant envelope infall or tidal interactions with a massive (external) planetary companion.  (d) A cartoon highlighting two representative outcomes for the evolution of icy aggregates as they migrate across a volatile condensation front.  The top behavior shows the case where ice loss to sublimation enhances $v_{\rm frag}$, and thereby promotes growth and drift; the bottom behavior is the opposite, resulting in a pileup of small, bare grains.   
}
\label{fig:sub_cartoon}
\end{figure}

The physical conditions in disks are subject to various (magneto-)hydrodynamic flows and instabilities that locally perturb gas pressures.  For example, the mechanics of disk dispersal itself can substantially reshape the disk structure.  Depending on the mass-loss profile, simulations of outflows from MHD-driven winds \citep{suzuki16,takahashi18} or photoevaporative flows \citep{alexander14,ercolano17} predict a ring-shaped pressure maximum at $\sim$tens of au, with a depleted (or even empty) cavity interior to it (see {\bf Figure \ref{fig:sub_cartoon}a}).

Even without imposing a special evolutionary state, generic fluid mechanics properties likely also play roles in substructure formation.  Turbulence generates stochastic $P$ modulations that concentrate particles and diminish their drift rates \citep{cuzzi01,pan11}.  That behavior might predict a disk mottled with substructures on the eddy scale ($\lesssim$\,$H_p$), but many simulations demonstrate that MHD turbulence tends to self-organize into more coherent features in the $P$ distribution, including spirals \citep{heinemann09b,flock11} and axisymmetric undulations \citep{johansen09,dittrich13}.  The latter are produced when the gas dynamics are modified by spontaneous, concentric concentrations of magnetic flux, which repel gas from regions of peak magnetic stress and pile it up at neighboring annuli \citep{uribe11,bai14b,simon14,bethune16,suriano17,suriano18}.  These zonal flows create narrow ($\Delta r \approx$\,few to 10\,$H_p$) depletions (gaps) and enhancements (rings) in $P(r)$ at $r \approx$\,tens of au that are expected to trap and concentrate solids (see {\bf Figure \ref{fig:sub_cartoon}b}).

In very dense regions of the disk, the ionization rate can be diminished enough to stifle turbulence from the MRI \citep{gammie96}.  The radial variation of $\alpha_{\rm t}$ into such a ``dead zone" modifies the gas flow and can thereby produce a strong, axisymmetric maximum in $P(r)$ at the laminar/turbulent boundary (\citealt{regaly12,dzyurkevich13}).  These transitions at dead zone boundaries can generate vortices through the Rossby wave instability \citep{lovelace99,lyra09}, resulting in radially narrow ($\Delta r \approx H_p$) but azimuthally extended ($\Delta \theta \gtrsim \pi/2$) pressure maxima \citep{lyra13,baruteau16}.  There are alternative ways to cultivate vortices, including baroclinic instabilities \citep{klahr03}, which can be amplified by feedback from the solids \citep{loren-aguilar15,loren-aguilar16}, and the vertical shear instability \citep{richard16}.  Vortices attract and concentrate migrating solids, making them especially compelling sites for planetesimal formation \citep{barge95,klahr97,klahr06}, and can also imprint long-lasting rings and gaps in $\Sigma_{\rm s}$ \citep[e.g.,][]{surville16}.  

For sufficiently dense and cold configurations, self-gravity can drive a global gravitational instability (GI) that imposes a large-scale spiral pattern onto the disk structure (see {\bf Figure \ref{fig:sub_cartoon}c}; \citealt{toomre64,boss97}).  A more unstable disk produces lower order modes (fewer arms) and more open (loosely-wrapped) patterns \citep[e.g.,][]{kratter16}.  The pressure peaks of the spiral waves concentrate and foster the growth of drifting particles \citep{rice04,dipierro15}.  At early evolutionary stages, asymmetric envelope accretion could drive the global GI \citep{laughlin94,tomida17,hall19}.  That infall process could also create vortices \citep{bae15}, generate an unstable shock that propagates in spiral density waves \citep{lesur15}, or magnetically imprint over-densities in concentric rings \citep{suriano17}.          

Various other modes of gas--particle coupling could also precipitate substructures in $\Sigma_{\rm s}$ and perhaps accelerate planetesimal formation if the solids-to-gas ratio is enhanced.  Two interesting examples are cases where ring-shaped particle over-densities are self-induced by a dynamical feedback (solids on gas) process \citep{drazkowska16,gonzalez17} or a viscous feedback instability where solid enhancements diminish $\alpha_{\rm t}$ and perturb the gas dynamics \citep{wunsch05,dullemond18}.  A special focus has been on the secular GI, which occurs when gas drag slows the self-gravitational collapse of solids enough to shear out the over-densities into narrow rings \citep{shariff11,youdin11}.  Simulations of the secular GI find $\sim$\,$H_p$-scale (perhaps clumpy) concentric peaks in $\Sigma_{\rm s}$, provided the turbulence is low ($\alpha_{\rm t} \lesssim 10^{-3}$; \citealt{takahashi14,takahashi16}).          

Obviously a remarkable variety of mechanisms in the broader fluid dynamics category can theoretically generate perturbations in $P$ (or $\Sigma_{\rm s}$) that are sufficient to mitigate the drift problem and promote the local concentration of solids.  In this general picture, the disk substructures produced by these internal, naturally-occurring mechanisms represent the fundamental {\it initial conditions} for planetesimal (and thereby planet) formation.

\subsubsection{Dynamical Interactions with Companions} \label{sec:pdx}
Gravitational (tidal) perturbations by a companion are a less subtle means of modifying disk properties, but are capable of generating a similar diversity of substructures as in the fluid mechanics category.  The emphasis here is on planets, although analogous effects are relevant for stellar binaries (Section \ref{sec:env}).  Once it has accumulated sufficient mass, a planetary companion interacts with the disk, generating spiral shocks that transfer angular momentum and repel disk material away from its orbit \citep{lin79,lin86,goldreich80}.  That perturbation can clear an annular gap in $\Sigma_{\rm g}$, with a width and depth that depend on the planet mass and the local turbulent diffusion and thermal structure of the gas disk \citep{kley12}.  The pressure maxima produced outside the gap, around the planetary orbit, can trap drifting solids \citep{rice06,paardekooper06,zhu12}.        

More dramatic perturbations to disk structures are produced by more massive (giant) planetary companions ($\gtrsim$\,M$_{\rm Jup}$).  Systems of multiple giant planets can have overlapping gaps that deplete $\Sigma_{\rm g}$ over a wide radial range (see {\bf Figure \ref{fig:sub_cartoon}a}; \citealt{dodson-robinson11,zhu11}), and may excite vortices or eccentric modes that generate strong azimuthal asymmetries in the pressure structure \citep{kley06,ataiee13,zhu14}.  If the companion orbit is inclined with respect to the disk plane, it can warp \citep{nealon18} or even dynamically isolate (``break") the disk into components with very different orientations \citep{owen17,zhu19b}.  Much lower mass planets ($\gtrsim$\,M$_\oplus$) can still generate substructures in $\Sigma_{\rm s}$, even if they only weakly perturb $P$ (see {\bf Figure \ref{fig:sub_cartoon}b}; \citealt{dipierro16,rosotti16}).  If turbulence is suppressed enough that the disk is essentially inviscid, low-mass planets can make a distinctive {\sf W}-shaped radial variation in the particle distribution (with the planet orbit at the central peak; \citealt{dong17}) and drive secondary and tertiary spiral arms that deposit angular momentum far interior to the planet orbit in near-circular shocks that also perturb $\Sigma_{\rm s}$ \citep{bae17,bae18a}.  Giant planets at large disk radii can also foster large-scale $m$=2 spiral modes interior to their orbits ({\bf Figure \ref{fig:sub_cartoon}c}; \citealt{zhu15,dong15}).

\subsubsection{Condensation Fronts} \label{sec:snowlines}
Substructures in $\Sigma_{\rm s}$ can also be induced {\it without} local pressure maxima (see Section \ref{sec:fluids}).  A popular example is associated with the sublimation of icy particles as they migrate across condensation fronts (snowlines).  In that scenario, three factors are relevant to consider.  First, ice sublimation is a net mass loss for the solids, and therefore $\Sigma_{\rm s}$ is depleted inside a snowline \citep{stammler17} over a radial range that depends on the coagulation timescales; efficient growth implies a narrow range.  Second, gas that has been liberated from ices can be mixed back across the snowline and re-condensed \citep{stevenson88,cuzzi04,ros13,ros19}.  This might help enhance particle growth and therefore $\Sigma_{\rm s}$ in a zone outside the snowline, with a width that depends on diffusion and migration rates.  

The third factor is perhaps most significant.  Ices can change the effective particle strengths (parameterized by the critical velocity for fragmentation, $v_{\rm frag}$), and thereby affect collision outcomes.  \citet{pinilla17} considered how molecular bonds in ices affect particle strengths, arguing that $v_{\rm frag}$ increases at the (CO or) CO$_2$ and NH$_3$ snowlines and decreases at the H$_2$O snowline.  This results in enhancements (depletions) of larger (smaller) particles between the H$_2$O and (CO or) CO$_2$ snowlines, although diffusion ($\alpha_{\rm t}$) affects the details (analogous to the top part of {\bf Figure \ref{fig:sub_cartoon}d}).  If $v_{\rm frag}$ decreases across a snowline, collisions can become destructive and the smaller fragments drift more slowly; the associated congestion increases $\Sigma_{\rm s}$ like a traffic jam \citep{birnstiel10,saito11}.  \citet{okuzumi16} argued that sintering during coagulation \citep{sirono11} diminishes $v_{\rm frag}$, and therefore can enhance $\Sigma_{\rm s}$ due to the reduced migration rates of small fragments, in narrow zones outside the snowlines of even rare volatile species.  These latter scenarios are illustrated in the bottom part of {\bf Figure \ref{fig:sub_cartoon}d}, although the sintering case would be shifted beyond the snowline (i.e., to the right in that diagram).  

The interplay between these factors is complex \citep[e.g.,][]{ciesla06,estrada16}, and likely complicated further by feedback reactions where particle accumulations affect the gas or solid dynamics \citep{drazkowska17,schoonenberg17,garate19}.  The potential outcomes are diverse.  However, the fundamental link to the disk temperatures means that the substructures associated with these mechanisms occur at special locations, and should be concentric and symmetric (presuming $T$ is dominated by a radial gradient).  In some sense, these limits to the flexibility of predictions from this idea could be helpful for observational tests, at least compared to the broad universe of options available from the fluid dynamics effects or planet-disk interactions outlined above.

\subsection{Signatures of Substructures}

Given the myriad physical processes capable of perturbing $P$ and/or $\Sigma_{\rm s}$, it is reasonable to expect that any given disk is riddled with substructures.  To find them and assess their origins, demographic dependences, and general roles in disk evolution and planet formation, measurements that characterize the forms, locations, sizes, and amplitudes of those features are crucial.  Some generic predictions about the properties of substructures can serve as useful guides for designing observations.  A stable perturbation to $P$ can have a characteristic size as small as $H_p$ \citep[e.g.,][]{dsharp6}.  For a standard disk temperature profile, $H_p/r \approx 0.05$--0.10.  That implies substructures might subtend only $\sim$5--50 milliarcseconds for projected separations of $\sim$0.05--0.5 arcseconds from the host star of a typical disk target ($d \approx 150$ pc).  Short-lived, stochastic features in the gas and the spatial concentrations of solids embedded in local pressure maxima could be even smaller.  Pressure perturbations with $\gtrsim$\,20\%\ amplitudes could be sufficient to trap drifting solids \citep{pinilla12b}.  That sets a crude benchmark on the sensitivity goals, although a focus on the strongly amplified signal from trapped solids can substantially improve the search yields.  

Obviously, those predictions foreshadow a challenging observational task.  However, the high-fidelity datasets at novel resolutions that have become available over the past five years have enabled a first detailed look at disk substructures.  The next sections explore some general properties and physical insights about such features from these initial measurements.

\subsubsection{Morphology}
\begin{figure}[t!]
\includegraphics[width=\textwidth]{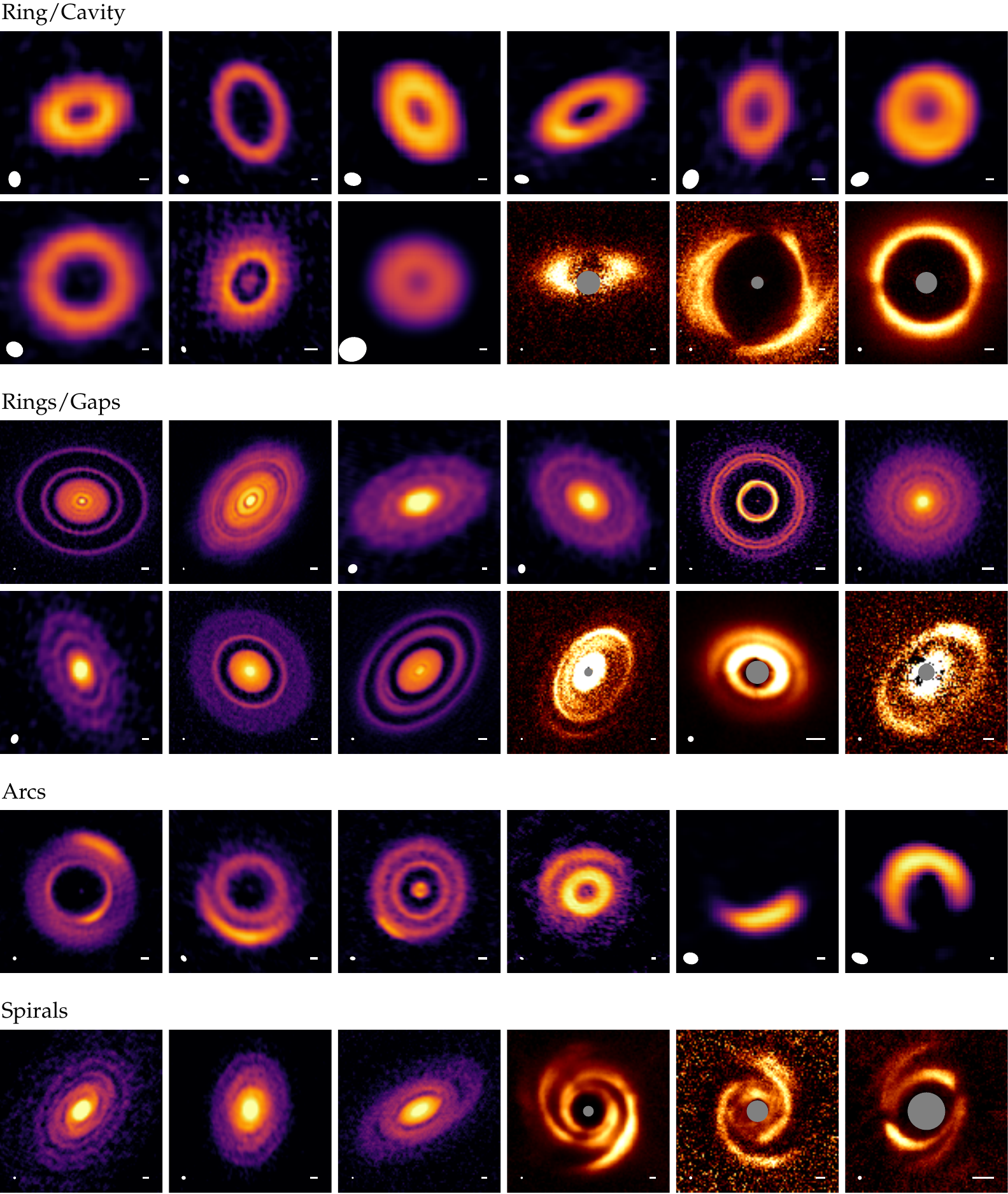}
\caption{{\it see following page.}}
\label{fig:morph}
\end{figure}

High resolution images of disks in their optical/infrared scattered light or mm continuum emission have identified substructures and characterized their morphologies at effectively all spatial scales down to the current resolution limits, $\sim$1--5 au (\citealt{zhang16,garufi18,dsharp1,long18}).  {\bf Figure \ref{fig:morph}} shows a gallery of images that highlights their morphological diversity.  While there are some subtle complexities, and a considerable overlap that indicates more of a continuum of substructure patterns, it is reasonable to group the morphological types into four broad categories.

\addtocounter{figure}{-1}
\begin{figure}[t!]
\caption{A gallery of disk substructure morphologies; the color maps for mm continuum and infrared scattered light match those in {\bf Figure \ref{fig:ims}}.  Resolutions are marked with white ellipses in the lower left corners of each panel; 10 au scalebars are shown in the lower right corners.  The mm continuum images are shown with an asinh stretch to $\sim$90\%\ of their peaks.  The scattered light images use a linear stretch, sometimes including an $r^2$ scaling, with the intent of matching the dynamic range shown in the literature: gray circles mark their coronagraphic spots (or bad pixel regions).    Ring/Cavity images, from left to right: (top) CIDA 9 \citep{long18}, Sz 91 \citep{tsukagoshi19b}, SR 24 S \citep{pinilla17c}, HD 34282 \citep{vanderplas17b}, IP Tau \citep{long18}, SR 21 \citep{vandermarel18}; (bottom) RX J1604.3-2130 \citep{pinilla18}, DM Tau \citep{kudo18}, DoAr 44 \citep{pinilla18}, IRS 48 \citep{follette15}, HD 142527 \citep{avenhaus14}, and RX J1604.3-2130 \citep{pinilla18c}.  Rings/Gaps images: (top) AS 209 \citep{dsharp8}, HL Tau \citep{brogan15}, V1094 Sco \citep{vanterwisga18}, DL Tau \citep{long18}, HD 169142 \citep{sperez19b}, RU Lup \citep{dsharp1}; (bottom) GO Tau \citep{long18}, Elias 24 \citep{dsharp1}, RX J1852.3-3700 \citep{villenave19}, RX J1615.3-3255 \citep{avenhaus18}, V4046 Sgr \citep{avenhaus18}, and HD 163296 \citep{monnier17}.  Arcs images: MWC 758 \citep{dong18}, SAO 206462 \citep{cazzoletti18}, HD 143006 \citep{dsharp10}, HD 163296 \citep{dsharp9}, V1247 Ori \citep{kraus17}, HD 142527 \citep{casassus13}.  Spirals images: IM Lup, WaOph 6, Elias 27 (all from \citealt{dsharp3}), SAO 206462 \citep{stolker17}, MWC 758 \citep{benisty15}, and HD 100453 \citep{benisty17}.  }
\end{figure}

\vspace{0.2cm}
\textbf{\textit{Ring/Cavity.}} This is the canonical morphology for a ``transition" disk \citep{espaillat14}, with a primary narrow ring (usually peaking at $r \approx {\rm tens}$ of au, though that is likely a selection bias) that encircles a depleted cavity \citep{pietu07,brown09,andrews11}.  The cavities are usually, but not always, cleared enough to depress the infrared SED \citep{calvet02,espaillat07,vandermarel16b}.  Disks with this morphology offer the best observational evidence for particle traps in local $P$ maxima, given their amplified (narrow, bright) concentrations of mm/cm continuum emission \citep{pinilla12b,pinilla18} and the lower amplitude and more spatially extended perturbations (both inside the cavity and to larger $r$) of line emission (tracing gas; \citealt{hughes09,vandermarel15,vandermarel16}) and scattered light (probing small grains that are well coupled to the gas; \citealt{dong12,mayama12,villenave19}).  

\vspace{0.1cm}
\textbf{\textit{Rings/Gaps.}} This refers to a concentric, axisymmetric pattern of alternating intensity enhancements (rings) and depletions (gaps).  It is the most common substructure morphology identified in the mm continuum \citep{brogan15,andrews16,dsharp2,long18} and scattered light \citep{quanz13,akiyama15,rapson15,ginski16,deboer16,avenhaus18}.  There are clear variants within this category, with options spanning from cases where the entire disk can be decomposed into narrow gaps and rings (e.g., CI Tau, \citealt{clarke18}; AS 209, \citealt{dsharp8}) to a single gap that separates an inner emission core from an outer ring (e.g., V883 Ori, \citealt{cieza16}; DS Tau, \citealt{long18}).

\vspace{0.1cm}
\textbf{\textit{Arcs.}}  Non-axisymmetric substructures seem to be rare, although some disks exhibit arc features that span a limited range of azimuth.  These arcs can be manifested as a partial ring around a central cavity, where the brightness asymmetry can range from severe ($\gtrsim$\,100$\times$; \citealt{vandermarel13,casassus13}) to mild ($\sim$2$\times$; \citealt{isella13,perez14,loomis17}).  Or, they can appear as {\it additional} substructures, located exterior to a ring/cavity morphology \citep{marino15,vandermarel16c,kraus17,boehler18} or in a gap \citep{dsharp9,dsharp10}.  

\vspace{0.1cm}
\textbf{\textit{Spirals.}}  Large-scale spiral patterns are most prevalent in scattered light images, ranging from pronounced $m$=2 modes with modest brightness asymmetries and a relatively open morphology \citep{muto12,grady13,akiyama16a} to more intricate, tightly-wrapped, and asymmetric structures \citep{hashimoto11,avenhaus14,garufi16,monnier19}.  There are only three known examples of disks around single star hosts that exhibit a spiral pattern in the mm continuum \citep{perez16,dsharp3}, in each case with a large, symmetric, two-armed pattern.  Extended, complex spirals have been identified in spectral line emission in three (different) cases \citep{tang12,tang17,christiaens14,teague19}.         

\vspace{0.1cm}
Of course, individual disks can include features with multiple morphological types.  Some examples were mentioned above for arc shapes, but there are also cases where rings, gaps, and a cavity (e.g., DM Tau, \citealt{kudo18}; Sz 129, \citealt{dsharp2}) or rings, gaps, and spirals \citep{dsharp3} co-exist in the same disk and for the same observational tracer.  Moreover, that mixing of morphological types can be striking when comparing images of the same disk in different tracers: the MWC 758 and SAO 206462 disks are favorite examples, showing spirals in scattered light \citep{benisty15,stolker16}, but rings, gaps, cavities, and arcs in the mm continuum \citep{dong18,cazzoletti18}.  This could be the hallmark of multiple mechanisms operating simultaneously, with different processes manifesting more clearly in tracers of the gas or the solids, or it could be indicative of a changing morphology as a function of altitude.  There are some tentative preferences for certain morphological types as a function of $M_{\ast}$, age, and perhaps other demographic properties \citep{garufi18}, but selection effects are still a considerable problem; an unbiased census for substructures is not yet available.

\subsubsection{Locations, Sizes, and Amplitudes}
The ring/gap substructures in disks are found at essentially any radial location, from the resolution limit (a few au) out to the detection threshold ($\lesssim$\,300 au; \citealt{dsharp2,vanterwisga18}).  Ring/cavity and arc substructures are preferentially identified at larger $r$ ($\sim$tens of au), although that is likely a resolution bias \citep{dsharp2,pinilla18b}.  There are no obvious relationships between substructure locations and host properties \citep{pinilla18b,long18,dsharp2,vandermarel19}.  There is a propensity to find more distant substructures in larger disks, but it is not clear if this is a physical connection (those disks are larger and brighter {\it because} they produced more distant substructures) or a trivial artifact.  First, more distant substructures should be larger (higher $H_p$) and therefore easier to find.  And second, even if substructures exist at the same large $r$ for disks with smaller $R_{\rm mm}$, there is no continuum emission at $r \gg R_{\rm mm}$ to trace them; they would likely be missed.  With those considerations in mind, the current suite of observations are consistent with the idea that there are no special locations essential for substructure formation.

The observed radial separations between neighboring ring or gap substructures in the mm continuum (peak-to-peak or trough-to-trough) span the range $\Delta r / \bar{r} \approx 0.2$--0.5 (where $\bar{r}$ is the midpoint between two features; \citealt{dsharp2}), corresponding to characteristic spacings $\Delta r \approx 2$--10\,$H_p$.  The low end of that range is uncertain, since more compact spacings (especially in the inner disk) can be missed with limited resolution.  A few disks include substructure pairings that have spacings commensurate with mean-motion resonances \citep{dsharp2}.  The two-dimensional nature of spiral substructures makes it difficult to define spacings (or locations).  The pitch angles of two-armed spirals are typically $\sim$10--20$^{\circ}$ (perhaps with modest radial gradients) in both scattered light \citep{muto12,yu19} or mm continuum \citep{dsharp3} measurements.  Simulations predict that the scattered light spirals should be more open than the mm continuum for a given disk, due primarily to the vertical temperature gradient (the scattered light tracers a higher, and therefore warmer, layer; \citealt{juhasz18}).  It is possible that more tightly-wound configurations could be mistaken for rings or arcs.       

Measurements of substructure widths are difficult because most features are not well resolved.  The broader rings/gaps in the mm continuum have ${\rm FWHM}/r \approx 0.1$--0.5, implying widths  $\approx H_p$ \citep{dsharp2,dsharp6,long18}.  The radial widths of most gaps, arcs, and spirals (i.e., in cuts perpendicular to tangent points) are $\lesssim$\,5--10 au.  The characteristic size distribution of substructures clearly extends below current resolution limits, but inferences of sub-resolution feature widths should be regarded with a healthy skepticism since they are strongly dependent on the adopted model prescription.  There is more diversity in the widths of rings, since there are many instances of extended (apparently smooth) cores, bands, or belts of emission that strain the simplistic morphological definition.  There is also considerable variety in the azimuthal extents of arcs, ranging from $\sim$5 to $\gtrsim$\,100$^\circ$ \citep[e.g.,][]{casassus13,tsukagoshi19}.  

Limited resolution also makes it challenging to quantify substructure contrasts.  If the features are not well resolved, peaks could be higher and troughs could be lower than they appear: contrasts should be considered lower bounds.  Moreover, mm continuum ring intensities will saturate at high optical depths: in that scenario, inferences of $\Sigma_{\rm s}$ contrasts could be much higher than the $I_\nu$ contrasts imply.  For the well-resolved ring/gap pairs, contrasts range from a few percent to a factor of $\sim$100 \citep{dsharp2,long18,avenhaus18}.  Contrasts inferred for ring/cavity or arc substructures can be even higher \citep{andrews11,vandermarel13,pinilla18}.  The contrasts between spirals and the local inter-arm material depends strongly on the tracer: while it can be high in scattered light \citep[e.g.,][]{benisty15,stolker16}, it is $\lesssim$\,3 in the few examples available for the mm continuum \citep{dsharp3}.      

There are fewer quantitative constraints available for the substructure properties based on the intensity distributions of key gas tracers.  In the ring/cavity cases, the gas is found to extend to smaller radii than the mm/cm solids, although it ultimately is depleted to comparable levels \citep{vandermarel15,vandermarel16}.  In some cases, spectral line measurements demonstrate that additional rings/gaps continue at distances well beyond the continuum emission \citep{huang18,dsharp8}.  Some preliminary analyses indicate that spectral line depletions (gaps) track their continuum counterparts \citep[e.g.,][]{isella16,dsharp9}.  But generally, the relative lack of emission line constraints is largely a technical limitation, since such observations at very high resolution are considerably more expensive than for the mm continuum or scattered light.  Such measurements are of high value, and should play increasingly important roles in future work.

\subsubsection{Optical Depth Fine-Tuning}
One subtle, puzzling outcome from the suite of new high resolution mm continuum observations potentially has wide-reaching implications.  Presuming that continuum emission is optically thin, and adopting a simple prescription for $T(r)$, the observed intensities at most substructure peaks imply  $\tau_\nu \approx 0.5$ (within a factor of two) at $\lambda = 1.3$ mm \citep{dsharp2,dsharp6}.  That value is suspiciously similar to the $\langle \tau_\nu \rangle$ estimated from the shape of the $R_{\rm mm}$--$L_{\rm mm}$ relation (Section \ref{sec:sizes}; \citealt{tripathi17}).  This fine-tuning {\it seems} artificial, as if it points to some important underlying process or flawed assumption.  So far, three ideas have been proposed to explain it.  

The first idea is a geometric argument: it presumes the emission is optically thick, but that the true brightness distribution is concentrated on sub-resolution scales \citep{tripathi17,andrews18}.  The second idea is a radiative transfer argument: it again presumes the emission is optically thick, but adds that self-scattering for particles with very high albedos ($\omega_\nu \gtrsim 0.9$) is important \citep{zhu19}.  In that case, scattering diminishes the intensities $I_\nu$, making the emission appear (marginally) optically thin even if the true $\tau_\nu$ is arbitrarily high.  That said, the high $\omega_\nu$ requirement maps onto a narrow $a_{\rm max}$ range ($\approx \lambda / 2\pi$): it is worth considering the plausibility of populating the $\tau_\nu \sim 1$ layers of disks with a very specific size distribution, or if this is just a change of variable for the fine-tuning problem (from $\tau_\nu$ to $a_{\rm max}$ at a special altitude).  The third idea is a physical argument: it takes the marginally thin optical depths at face value and considers a self-regulating exchange between $\kappa_\nu$ and $\Sigma_{\rm s}$ in local pressure maxima \citep{stammler19}.  In this case, particle evolution simulations can explain the data if a fraction ($\sim$10\%) of the solid mass is converted into planetesimals whenever the solids-to-gas ratio approaches unity.  

While these ideas are fleshed out, it is important to evaluate more generally the implications of the two optically thick hypotheses.  High optical depths are natural explanations for the $R_{\rm mm}$--$L_{\rm mm}$ relation, as well as the low $\alpha_{\rm mm}$ \citep{zhu19,liu19} or $\varepsilon$ in the inner disk regions ({\bf Figure \ref{fig:mm_indices}}) and substructure rings \citep{carrasco-gonzalez16,carrasco-gonzalez19,tsukagoshi16,huang18}.  However, this would force a re-evaluation of traditional estimates of disk masses and densities (Section \ref{sec:structure}), shift the interpretations of various demographic trends (Section \ref{sec:demographics}), and perhaps severely complicate the analyses of spectral line observations \citep[e.g.,][]{weaver18}.  Given the scope of what is at stake, identifying the origins of this fine-tuning puzzle is certainly a high priority.

\subsubsection{Sample Bias and Resolution Limitations}
The current sample of very high resolution measurements in the mm continuum or scattered light is biased in favor of larger, brighter disks that preferentially orbit more massive host stars \citep{dsharp1,garufi18}.  This is a practical restriction, but also in some sense by design.  To find and measure substructures with characteristic sizes $\approx H_p$ at the resolutions (and inner working angles) available from current facilities requires them to be located at $r \gtrsim 20$--50 au.  A more representative target, around a host with half the mass and 5--10$\times$ lower $L_{\rm mm}$, would typically have its tracer emission concentrated within that critical radius.  If the substructures closer to the stellar hosts are like those we find at larger $r$ (i.e., widths $\lesssim H_p$), these smaller disks will {\it appear smooth} even if they are also riddled with substructures.          

Indeed, there is plenty of evidence that the distribution of continuum substructure sizes extends below $H_p$-scales and current resolution limits (e.g., from the many partially resolved features mentioned by \citealt{dsharp2}).  The advantageous distance of the TW Hya disk offers an instructive example: nearly all of its rings and gaps are narrow enough that they would be indistinguishable from an unperturbed emission profile if the system were moved out to the nearest young clusters \citep[see][]{andrews16,huang18}.  This is not to say that efforts to find especially large substructures in more representative samples are undesirable: rather, the point is that any conclusions about the prevalence of substructures need to be contextualized to the accessible range of substructure sizes.

\subsubsection{Kinematics}
Another approach to quantifying substructures in $P(r)$ involves a search for the associated deviations in the radial velocity profile of the gas, $v_\theta$.  These non-Keplerian motions can be identified when the residual velocity profile, $\delta v_\theta = (v_\theta - v_{\rm kep}) / v_{\rm kep}$, flips sign over a narrow spatial range: that signal can then be related to the local pressure gradient (Section \ref{sec:fix_drift}).  Observational constraints on the spatial patterns of $\delta v_\theta$ offer independent, kinematic insights on substructures that complement their intensity characterizations based on high resolution spectral line or continuum images.

With sufficient resolution and sensitivity, (sub-)mm spectral line datasets can be used to reconstruct azimuthally-averaged $\delta v_\theta(r)$ profiles in the line photosphere layers \citep{teague18a,teague18b,sperez18}.  That technique has been used to identify kinematic perturbations with $\sim$5--10\%\ amplitudes (at roughly percent level precision) that spatially coincide with known substructures in the mm continuum.  Naturally, the azimuthal averaging required to tease out small $\delta v_\theta$ signals means that it is unclear whether or not those kinematic deviations are axisymmetric.  If the perturbations are tracing the spiral wakes expected from planet-disk interactions, the $\delta v_\theta$ signal would peak in the immediate vicinity of the perturber \citep{sperez15,sperez18,kanagawa15,teague18a,dsharp7}.  Especially strong localized $\delta v_\theta$ sign-flips have been identified in a few cases \citep{pinte18b,pinte19,casassus19}, lending striking support to such a dynamical origin even if the perturber itself cannot yet be directly detected.

Such kinematics constraints on disk properties are still in a relatively early stage of development, both observationally and theoretically.  Nevertheless, they hold immense promise in their synergy with the more traditional measurements emphasized throughout this review.  Leveraging these techniques together will be a necessary step in the push toward a more quantitative characterization of substructure properties.

\subsubsection{Vertical Perturbations}
\begin{figure}[t!]
\includegraphics[width=\textwidth]{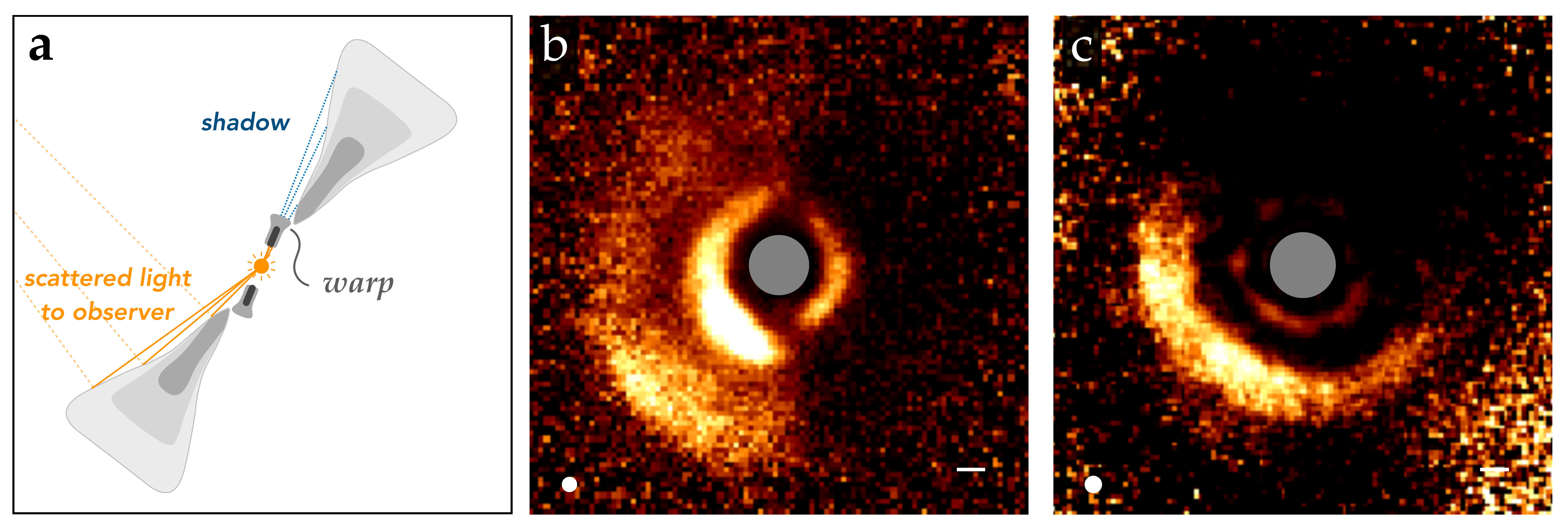}
\caption{(a) A schematic illustration of how a disk warp can induce shadowing in scattered light images (see also \citealt{marino15}).  (b) Two examples of shadowing from large-scale asymmetries in infrared scattered light images, from the disks around (b) HD 143006 \citep{benisty18} and (c) WRAY 15- 788 \citep{bohn19}.  Image annotations are as in {\bf Figure \ref{fig:morph}}.  A stronger warp, or even a ``broken" disk geometry can change the azimuthal extent of the shadows (e.g., see the narrow shadows for the HD 100453 disk, at bottom right in {\bf Figure \ref{fig:morph}}; \citealt{benisty17}).  }
\label{fig:shadows}
\end{figure}

The discussion above emphasizes the ($r$, $\theta$)-plane, but there are also clear signs of substructures in the vertical ($z$) dimension.  These can be inferred indirectly from infrared variability  \citep{muzerolle09,flaherty12,rebull14}, especially in cases where the variations follow a ``see-saw" spectral pattern with enhancements at shorter $\lambda$ (lower, warmer $r$) accompanied by depletions at longer $\lambda$ (higher, cooler $r$), and vice versa \citep{espaillat11}.  These time-domain phenomena are presumably associated with the stellar obscuration and disk shadowing that occur when the vertical distribution of inner disk material is perturbed \citep[e.g.,][]{turner10}.  In some cases, that shadowing can be seen directly in scattered light images \citep{garufi14b,benisty18}, as illustrated in {\bf Figure \ref{fig:shadows}}.  This obscuration can be quite variable on short timescales, indicating a clumpy distribution of occulting material in the inner disk \citep[e.g.,][]{stolker17,pinilla18c}.  \citet{debes17} discovered an especially compelling example, where an outer disk shadow moves at a rate consistent with inner disk orbital timescales.  There is a hypothesis that spirals identified in scattered light could be associated with such behavior  \citep{kama16c,montesinos16}.              

In most examples, shadows require a persistent vertical substructure in the inner disk.  This is often associated with a warp, since even modest changes in the orbital inclination distribution of the disk material generate pronounced observational effects \citep{nealon19}.  Warped geometries have been inferred kinematically with resolved spectral line data, based on the spatial variation of the projected (line-of-sight) velocities \citep{rosenfeld12a,rosenfeld14,casassus15,loomis17}.  Large warps or misalignments (``broken" disks) can also induce scattered light shadows at larger $r$; the locations and azimuthal extents of the shadows help constrain the inner disk geometry \citep{marino15,stolker16,facchini18,pinilla18c,casassus18}.  

In any case, a variety of vertical substructures are inferred at locations (and with sizes) that are well below current resolution capabilities.  Nevertheless, their effects are manifested on much larger scales in the ($r$, $\theta$) behavior of key tracers.  The important lesson is that current observations are sensitive to substructures in all three spatial dimensions.

\subsection{Emerging Insights}

The specific topic of disk substructures has generated immense interest.  The many new opportunities for high resolution observations of disks have triggered a marked pivot in the field toward their interpretation.  Despite the deluge in the literature, it is worth keeping in mind that assessments of the broader impacts that substructures have on disk evolution and planet formation are still being actively developed.  Nevertheless, it is also clear from their prevalence alone that substructures are {\it fundamental} aspects of disks: they likely have profound effects on every practical and physical facet of planet formation research.  

\begin{figure}[t!]
\includegraphics[width=\textwidth]{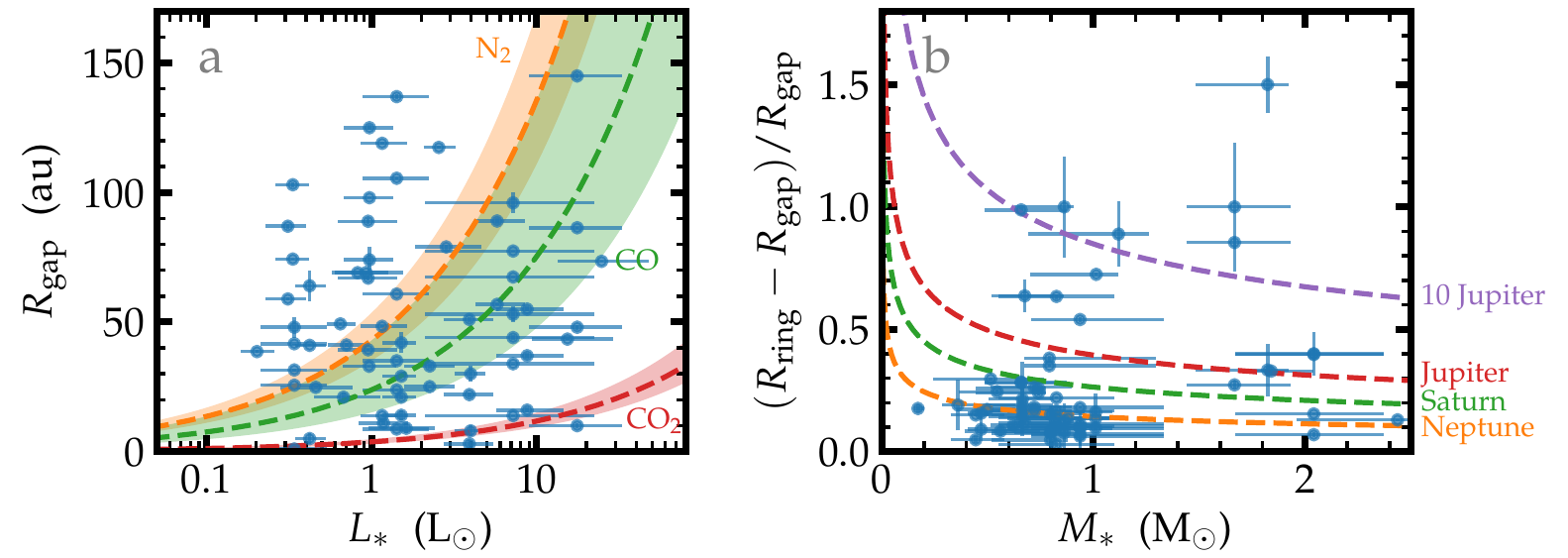}
\caption{(a) The locations of gap substructures as a function of $L_\ast$ (data from \citealt{dsharp2,long18}).  The overlaid shaded regions mark the expected locations for the condensation fronts of abundant volatiles (as labeled, following \citealt{dsharp2}).  There is no clear pattern indicating a connection between substructure locations and special disk temperatures.  (b) The fractional separations between ring and gap substructure pairs (from these same studies) as a function of $M_\ast$.  The overlaid curves mark the masses of planets that might be responsible for opening the gaps, following the simplified assumptions of \citet{long18} and \citet{lodato19}.  The implied masses (and orbits, examining panel a) probe a very different range of parameter-space from the mature exoplanet population \citep[see also][]{dsharp7}.  }
\label{fig:substruct}
\end{figure}

The current priority is to develop a more quantitative understanding of the physical mechanism(s) responsible for generating substructures and how they impact (or perhaps reveal ongoing) planet formation.  At this point, it is fair to conclude that none of the potential origins discussed in Section \ref{sec:origins} can be categorically excluded.  Without rehashing a detailed comparison of the data and model predictions, there are a few generic points about the emerging themes of this analysis that are worth highlighting.    

One important conclusion is that the simple empirical signatures expected from models of particle migration around snowlines (Section \ref{sec:snowlines}) are not observed \citep{vanterwisga18,long18,dsharp2,vandermarel19}.  Those signatures include a rough $L_\ast^{0.5}$ scaling to the pattern of ring/gap locations in the mm continuum and an $r^{-0.5}$ spacing between them, and originate from the hypothesis that such substructures occur at special locations corresponding to the condensation temperatures of abundant volatiles.  {\bf Figure \ref{fig:substruct}a} demonstrates that these patterns are not obvious in the data.  However, there is considerable diversity in the model predictions, and one cannot rule out that any individual feature (or the collection of features in individual disks) might be associated with snowlines \citep{zhang15,vandermarel18b}.  More extensive vetting of this hypothesis will consider the non-trivial uncertainties in the disk temperatures (Section \ref{sec:temp}) and the dependence of condensation temperatures on the bulk ice composition and local gas pressure.  Meanwhile, some theoretical consensus would be useful: recent models predict either emission enhancements or depletions, located either inside or outside the snowline (Section \ref{sec:snowlines}).  It would help to know whether or not that fungible range of outcomes represents an inherent physical ambiguity.                  

There are additional insights that disfavor the snowline hypothesis, in that there is quantitative evidence that many observed substructures do trace particle traps at local gas pressure maxima.  The subset of resolved mm continuum rings are found to have the high amplitudes and narrow widths ($<$~$H_p$) predicted for these traps \citep{dsharp6}.  Complementary support for that conclusion is also available from the demographics \citep{pinilla18}, kinematics \citep{teague18a,teague18b}, diverse tracer-dependent morphologies in ring/cavity substructures \citep{dong12,vandermarel15}, and narrow azimuthal extents of continuum emission with low $\varepsilon$ at the peaks of arc substructures \citep{birnstiel13,vandermarel15b,casassus15b}.  These constraints lend credibility to the fluid mechanics or planet-disk interactions hypotheses for substructure origins, although robustly discriminating between those options is perhaps not yet practical \citep{flock15,ruge16,dong18c}.

That said, the mechanism(s) that generate these very detailed disk substructures really color the global perspective on what information can be gleaned from disk properties.  If fluid perturbations from various (M)HD processes are ultimately responsible, then disks are genuinely in a classical ``protoplanetary" phase, representative of incipient planet formation.  Detailed measurements of their properties would illustrate how disk substructures are fundamental for making planetesimals.  Much of the immediate progress to be made in testing this hypothesis will come from enhanced computational capabilities, to help develop more robust and discriminating predictions.  The alternative hypothesis, that substructures are instead produced by perturbations from already-formed planetary systems, has subtle but profound implications for the standard principles of planet formation theories.  

To discuss those implications, it helps to consider the masses and orbits of the youthful planetary systems that are inferred from the morphologies of the disk substructures.  With reference to simulations of disk-planet interactions, the locations, widths, and depths of disk gaps \citep{dsharp7,lodato19} and cavities \citep[e.g.,][]{zhu12} suggest perturber masses from a few M$_\oplus$ to $\sim$10 M$_{\rm Jup}$ orbiting at semimajor axes of $\sim$10--150 au (around a representative $\sim$Sun-like host).  {\bf Figure \ref{fig:substruct}b} illustrates some representative results.  Note that this is a different (complementary) region of parameter-space for planetary system architectures than has been probed in exoplanet surveys around mature host stars (early direct imaging constraints overlap at the high-mass end).  The timescales to form such planets in the standard (core accretion) formation theory are considerably longer than the typical system ages ($\sim$1--3 Myr).  So, if such planets already exist, they require that the formation process starts very early -- perhaps overlapping with the epoch of star (and disk) formation itself -- or is substantially accelerated (e.g., perhaps with some variant of pebble accretion; \citealt{ormel10,lambrechts12}).  In either case planetesimal formation must be efficient and prolific over a wide range of disk radii, presumably aided by an {\it earlier} generation of substructures (perhaps generated by an assortment of fluid dynamical mechanisms that are more prevalent during the embedded phase).      

The planet-disk interaction hypothesis is certainly an exciting prospect, since it offers potential opportunities to observationally constrain planet formation timescales, planetary accretion and satellite formation (through studies of circumplanetary material), and the evolution of planetary system architectures (i.e., planetary migration).  And there is increasing confidence in this option, especially from observations of strong gas depletion in disk cavities \citep{vandermarel15,vandermarel16}, spatially isolated perturbations to the disk gas dynamics that coincide with substructures \citep{casassus19,pinte19}, and most importantly the direct imaging detections of young giant planets in the cavity of the PDS 70 disk \citep{keppler18,muller18,haffert19}.  Nevertheless, there is still much work to do regarding the origins of disk substructures: assessing these forking paths is the single most important task of the coming decade in this field.

\section{SYNOPSIS \& OUTLOOK} \label{sec:summary}

\begin{summary}[SUMMARY POINTS]
\begin{enumerate}
\item New quantitative insights on key structure parameters are starting to bear fruit, but the intrinsic uncertainties on physical conditions suggest it is important to consider {\it empirical} metrics and translate predictions into the data-space.
\item There is clear evidence for multi-dimensional demographic relationships between disk properties (mm continuum luminosities, sizes) and various dynamical ($M_\ast$), environmental (e.g., multiplicity), and evolutionary factors.
\item While observations provide strong, qualitative support for the predicted behaviors of particle growth and migration models (particularly the spatial segregation of particle sizes and the radial gradient in the solids-to-gas ratio), there is a timescale discrepancy that suggests a smooth gas pressure profile is a poor assumption.
\item Small-scale substructures with a range of morphologies are found on scales comparable to $H_p$ throughout many (and perhaps all) disks, with dimensions and contrasts consistent with expectations for particle trapping in local gas pressure maxima. 
\item The origins of these substructures are not yet clear: leading contenders include an assortment of fluid instabilities or dynamical interactions with young planets.  In any case, the implications are profound, in that (physical and observational) disk properties could be determined by perturbations at very small scales.
\end{enumerate}
\end{summary}

\begin{issues}[FUTURE ISSUES]
\begin{enumerate}
\item Quantitative constraints on the density and temperature structure of the gas in disks are essential to enable progress in the field.  Investments in deep, high resolution observations of molecular spectral line emission should be prioritized.
\item The velocity dimension of those spectral line datasets is a rich frontier for reaping physical information.  Kinematic studies of turbulent motions and non-Keplerian deviations to the velocity field are expected to create many new opportunities to address theoretical predictions from a complementary perspective.  
\item There is vast potential to develop a better understanding of disk properties encoded in their demographic relationships (and associated scatter).  Expanding that work to more diverse samples (e.g., ages, multiplicity parameters) and additional metrics (e.g., deeper, resolved spectral line data) would be highly valuable.  
\item Progress on quantifying the evolution of disk solids will require a continued pursuit of spatial variations in the mm/cm continuum spectrum and polarization properties.  Folding those properties into demographics studies would be especially illuminating.
\item A shift to quantitative characterizations of disk substructures -- density contrasts, diffusion, particle sizes, kinematics, or their empirical equivalents -- would provide welcome guidance for theoretical predictions and help better assess their origins.
\item An appraisal of how the prevalence, morphologies, locations, and scales of disk substructures depend on host properties, global disk characteristics, environment, and especially age could reveal patterns that help contextualize general demographic trends and clarify the mechanics and variety of the processes that generate them.
\end{enumerate}
\end{issues}

\section*{DISCLOSURE STATEMENT}
The author is not aware of any affiliations, memberships, funding, or financial holdings that
might be perceived as affecting the objectivity of this review. 

\section*{ACKNOWLEDGMENTS}
I am very grateful to Myriam Benisty, Til Birnstiel, Marco Tazzari, and David Wilner for their detailed counsel and assistance with some of the ideas and figures in this review.  Xuening Bai, John Carpenter, Kees Dullemond, Michiel Hogerheijde, Feng Long, Antonella Natta, Karin {\"O}berg, Paola Pinilla, Richard Teague, Nienke van der Marel, Ewine van Dishoeck, Jonathan Williams, and Zhaohuan Zhu provided helpful, detailed comments on drafts that helped to substantially improve the writing, organization, and general presentation of ideas.  Juan Alcal{\'a}, Jane Huang, Meredith Hughes, Kevin Flaherty, and James Owen offered useful advice that helped crystallize ideas and writing strategies.  I also want to thank Megan Ansdell, Paolo Cazzoletti, Lucas Cieza, Ian Czekala, Ruobing Dong, Nathan Hendler, Stefan Kraus, Tomoyuki Kudo, Ryan Loomis, Sebastian P{\'e}rez, Dary Ruiz-Rodriguez, Pat Sheehan, Roy van Boekel, Gerrit van der Plas, and Sierk van Terwisga, in addition to some of those listed above, for kindly sharing their data.  This review made extensive use of the \textsf{numpy} \citep{vanderwalt11}, \textsf{matplotlib} \citep{hunter07}, \textsf{astropy} \citep{astropy18}, and \textsf{dsharp-opac} \citep{dsharp5} software packages, as well as NASA's Astrophysics Data System and the SIMBAD database, operated at CDS, Strasbourg, France.


\end{document}